\documentclass{article} 
\usepackage{iclr2021_conference,times}

\usepackage{amsmath,amsfonts,bm}

\def\figref#1{figure~\ref{#1}}

\def\eqref#1{equation~\ref{#1}}

\def\1{\bm{1}}

\def\mA{{\bm{A}}}
\def\mB{{\bm{B}}}
\def\mC{{\bm{C}}}
\def\mD{{\bm{D}}}

\def\mH{{\bm{H}}}

\def\mK{{\bm{K}}}

\def\mM{{\bm{M}}}

\def\mP{{\bm{P}}}
\def\mQ{{\bm{Q}}}

\def\mS{{\bm{S}}}

\def\mU{{\bm{U}}}

\def\mW{{\bm{W}}}
\def\mX{{\bm{X}}}
\def\mY{{\bm{Y}}}

\DeclareMathAlphabet{\mathsfit}{\encodingdefault}{\sfdefault}{m}{sl}
\SetMathAlphabet{\mathsfit}{bold}{\encodingdefault}{\sfdefault}{bx}{n}

\DeclareMathOperator{\Tr}{Tr}

\usepackage{tikz}
\usetikzlibrary{decorations.pathreplacing,decorations.pathmorphing}

\usepackage{algorithm}
\usepackage{algorithmic}
\floatname{algorithm}{Procedure}

\usepackage{hyperref}
\usepackage{url}
\usepackage{physics}
\usepackage{graphicx}
\usepackage{wrapfig} 
\usepackage{multirow}

\title{Quantum Deformed Neural Networks}

\author{Roberto Bondesan \& Max Welling \\
Qualcomm AI Research\thanks{Qualcomm AI Research is an initiative of Qualcomm Technologies, Inc.} \\
\texttt{\{rbondesa, mwelling\}@qti.qualcomm.com}\\
}

\iclrfinalcopy 
\begin{document}

\maketitle

\begin{abstract}
We develop a new quantum neural network layer designed to run efficiently on a quantum computer but that can be simulated on a classical computer when restricted in the way it entangles input states. We first ask how a classical neural network architecture, both fully connected or convolutional, can be executed on a quantum computer using quantum phase estimation. We then deform the classical layer into a quantum design which entangles activations and weights into quantum superpositions. While the full model would need the exponential speedups delivered by a quantum computer, a restricted class of designs represent interesting new classical network layers that still use quantum features. We show that these quantum deformed neural networks can be trained and executed on normal data such as images, and even classically deliver modest improvements over standard architectures.  \end{abstract}

\section{Introduction}

Quantum mechanics (QM) is the most accurate description for physical phenomena at very small scales, such as the behavior of molecules, atoms and subatomic particles. QM has a huge impact on our every day lives through technologies such as lasers, transistors (and thus microchips), superconductors and MRI. 

A recent view of QM has formulated it as a (Bayesian) statistical methodology that only describes our subjective view of the (quantum) world, and how we update that view in light of evidence (i.e.~measurements) \citep{hooft2016cellular,Fuchsetal}. This is in perfect analogy to the classical Bayesian view, a statistical paradigm extensively used in artificial intelligence where we maintain probabilities to represent our beliefs for events in the world.  

The philosophy of this paper will be to turn this argument on its head. If we can view QM as just another consistent statistical theory that happens to describe nature at small scales, then we can also use this theory to describe \emph{classical} signals by endowing them with a Hilbert space structure. In some sense, the 'only' difference with Bayesian statistics is that the positive probabilities are replaced with complex 'amplitudes'. This however has the dramatic effect that, unlike in classical statistics, interference between events now becomes a possibility. 
In this paper we show that this point of view uncovers new architectures and potential speedups for running neural networks on quantum computers.

We shall restrict our attention here to binary neural networks. 
We will introduce a new class of quantum neural networks and interpret them as generalizations of probabilistic binary  neural  networks, discussing  potential  speedups  by  running  the  models  on  a quantum computer. Then we will devise classically efficient algorithms to train the networks for a restricted set of quantum circuits.  We present results of classical simulations of the quantum neural networks on real world data sizes and related gains in accuracy due to the quantum deformations. Contrary to almost all other works on quantum deep learning, our quantum neural networks can be simulated for practical classical problems, such as images or sound. The quantum nature of our models is there to increase the flexibility of the model-class and add new operators to the toolbox of the deep learning researcher, some of which may only reach their full potential when quantum computing becomes ubiquitous. 

\subsection{Related work}

In \cite{farhi_neven} variational quantum circuits that can be learnt via stochastic gradient descent were introduced. Their performance could be studied only on small input tasks such as classifying $4\times 4$ images, due to the exponential memory requirement to simulate those circuits.
Other works on variational quantum circuits for neural networks are \cite{baqprop, beer}. Their focus is similarly on the implementation on near term quantum devices and these models cannot be efficiently run on a classical computer. Exceptions are models which use tensor network simulations \citep{qcnn,huggins} where the model can be scaled to $8\times 8$ image data with $2$ classes, at the price of constraining the geometry of the quantum circuit \citep{huggins}.
The quantum deformed neural networks introduced in this paper are instead a class of variational quantum circuits that can be scaled to the size of data that are used in traditional neural networks as we demonstrate in section \ref{sec:exp}.

Another line of work directly uses tensor networks as full precision machine learning models that can be scaled to the size of real data \citep{stoudenmire, Liu, convac, convac2}. 
However the constraints on the network geometry to allow for efficient contractions limit the expressivity and performance of the models. See however \cite{cheng2020supervised} for recent promising developments.  Further, the tensor networks studied in these works are not unitary maps and do not directly relate to implementations on quantum computers.

A large body of work in quantum machine learning focuses on using quantum computing to provide speedups to classical machine learning tasks \citep{Biamonte,ciliberto,quantumDL}, culminating in the discovery of quantum inspired speedups in classical algorithms \citep{tang}. 
In particular, \citep{allcock, quantum_neuron, Schuld_2015, kerenidis2019quantum} discuss quantum simulations of classical neural networks with the goal of improving the efficiency of classical models on a quantum computer. Our models differ from these works in two ways: i) we use quantum wave-functions to model weight uncertainty, in a way that is reminiscent of Bayesian models; ii) we design our network layers in a way that may only reach its full potential on a quantum computer due to exponential speedups, but at the same time can, for a restricted class of layer designs, be simulated on a classical computer and provide inspiration for new neural architectures. 
Finally, quantum methods for accelerating Bayesian inference have been discussed in \cite{quantumGP1,quantumGP2} but only for Gaussian processes while in this work we shall discuss relations to Bayesian neural networks.

\section{Generalized probabilistic binary neural networks}
\label{sec:generalization}

Binary neural networks are neural networks where both weights and activations are binary.
Let $\mathbb{B} = \{0, 1\}$. 
A fully connected binary neural network layer maps the $N_\ell$ activations $\bm{h}^{(\ell)}$ at level $\ell$ to the $N_{\ell+1}$ activations $\bm{h}^{(\ell+1)}$ at level $\ell+1$ using weights $\mW^{(\ell)}\in \mathbb{B}^{N_\ell N_{\ell+1}}$:
\begin{align}
    \label{eq:hell}
    h_j^{(\ell+1)} =
    f(\mW^{(\ell)}, \bm{h}^{(\ell)})
    =
    \tau\left(\frac{1}{N_\ell+1}\sum_{i=1}^{N_\ell} W^{(\ell)}_{j, i} h^{(\ell)}_i\right)\,, 
    \quad
    \tau(x) = 
    \begin{cases}
    0 & x < \frac{1}{2}\\
    1 & x \ge \frac{1}{2}
    \end{cases}
    \,.
\end{align}
We divide by $N_\ell+1$ since the sum can take the $N_\ell+1$ values $\{0,\dots,N_\ell\}$. We do not explicitly consider biases which can be introduced by fixing some activations to $1$.
In a classification model $\bm{h}^{(0)} = \bm{x}$ is the input and the last activation function is typically replaced by a softmax which produces output probabilities $p(\bm{y} |\bm{x}, \mW )$, where
$\mW$ denotes the collection of weights of the network. 

Given $M$ input/output pairs $\mX = (\bm{x}^1,\dots, \bm{x}^M), \mY = (\bm{y}^1,\dots,\bm{y}^M)$, a frequentist approach would determine the binary weights so that the likelihood $p(\mY |\mX, \mW )=\prod_{i=1}^M p(\bm{y}_i |\bm{x}_i, \mW )$ is maximized.
Here we consider discrete or quantized weights and take the approach of variational optimization \cite{staines2012variational}, which introduces a weight distribution $q_\theta(\mW)$ to devise a surrogate differential objective. For an objective $O(\mW)$, one has the bound $\max_{\mW\in\mathbb{B}^N} O(\mW) \ge \mathbb{E}_{q_\theta(\mW)}[O(\mW)]$, and the parameters of $q_\theta(\mW)$ are adjusted to maximize the lower bound. In our case we consider the objective:
\begin{align}
    \label{eq:obj_vo}
    \max_{\mW\in\mathbb{B}^N} \log p(\mY |\mX, \mW ) \ge 
    {\cal L}
    :=
    \mathbb{E}_{q_\theta(\mW)}[\log p(\mY |\mX, \mW )]
    =
    \sum_{i=1}^M
    \mathbb{E}_{q_\theta(\mW)}[\log p(\bm{y}_i |\bm{x}_i, \mW )]
    \,.
\end{align}
While the optimal solution to \eqref{eq:obj_vo} is a Dirac measure, one can add a regularization term ${\cal R}(\theta)$ to keep $q$ soft. In appendix \ref{sec:elbo_vo} we review the connection with Bayesian deep learning, where $q_\theta(\mW)$ is the approximate posterior, ${\cal R}(\theta)$ is the KL divergence between $q_\theta(\mW)$ and the prior over weights, and the objective is derived by maximizing the evidence lower bound.

In both variational Bayes and variational optimization frameworks for binary networks, we have a variational distribution $q(\mW)$ and probabilistic layers where
activations are random variables.
We consider an approximate posterior factorized over the layers: $q(\mW) = \prod_{\ell=1}^L q^{(\ell)}(\mW^{(\ell)})$.
If $\bm{h}^{(\ell)} \sim p^{(\ell)}$, \eqref{eq:hell} leads to the following recursive definition of distributions:
\begin{align}
    \label{eq:py}
    p^{(\ell+1)}(\bm{h}^{(\ell+1)}) 
    &= 
    \sum_{\bm{h}\in \mathbb{B}^{N_\ell}}
    \sum_{\mW\in \mathbb{B}^{N_\ell N_{\ell+1}}}
    \delta(\bm{h}^{(\ell+1)} - f(\mW^{(\ell)}, \bm{h}^{(\ell)}))
    p^{(\ell)}(\bm{h}^{(\ell)}) 
    q^{(\ell)}(\mW^{(\ell)})\,.
\end{align} 
We use the shorthand $p^{(\ell)}(\bm{h}^{(\ell)})$ for $p^{(\ell)}(\bm{h}^{(\ell)} | \bm{x})$ and the $\bm{x}$
dependence is understood.
The average appearing in \eqref{eq:obj_vo} can be written as an average over the network output distribution:
\begin{align}
    \mathbb{E}_{q_\theta(\mW)}[\log p(\bm{y}_i |\bm{x}_i, \mW )]
    =
    -\mathbb{E}_{p^{(L)}(\bm{h}^{(L)}) }
    [g_i(\bm{y}_i, \bm{h}^{(L)})]
    \,,
\end{align}
where the function $g_i$ is typically MSE for regression and cross-entropy for classification. 

In previous works \citep{shayer,pbnn}, the approximate posterior was taken to be factorized:
$q(\mW^{(\ell)}) = \prod_{ij} q_{i,j}(W^{(\ell)}_{i,j})$, which results in a factorized activation distribution as well:
$p^{(\ell)}(\bm{h}^{(\ell)}) = \prod_{i} p_{i}^{(\ell)}(h^{(\ell)}_{i})$.
\citep{shayer,pbnn} used the local reparameterization trick \cite{kingma2015variational} to sample activations at each layer. 

The quantum neural network we introduce below will naturally give a way to sample efficiently from complex distributions and in view of that we here generalize the setting: we act with a stochastic matrix $\mS_\phi(\bm{h}', \mW'| \bm{h}, \mW)$ which depends on parameters $\phi$ and correlates the weights and the input activations to a layer as follows: 
\begin{align}
    \pi_{\phi,\theta}(\bm{h}', \mW') &= 
    \sum_{\bm{h}\in \mathbb{B}^N}
    \sum_{\mW\in \mathbb{B}^{NM}} 
    \mS_\phi(\bm{h}', \mW'| \bm{h}, \mW)
    p(\bm{h}) 
    q_\theta(\mW)
    \,.
\end{align}
To avoid redundancy, we still take $q_\theta(\mW)$ to be factorized and let $\mS$ create correlation among the weights as well. 
The choice of $\mS$ will be related to the choice of a unitary matrix $\mD$ in the quantum circuit of the quantum neural network.
A layer is now made of the two operations, $\mS_\phi$ and the layer map $f$, resulting in the following output distribution:
\begin{align}
    \label{eq:py_S}
    p^{(\ell+1)}(\bm{h}^{(\ell+1)}) 
    =
    \sum_{\bm{h}\in \mathbb{B}^N_\ell}
    \sum_{\mW\in \mathbb{B}^{N_\ell N_{\ell+1}}}
    \delta(\bm{h}^{(\ell+1)} - f(\mW^{(\ell)}, \bm{h}^{(\ell)}))
    \pi^{(\ell)}_{\phi,\theta}(\bm{h}^{(\ell)}, \mW^{(\ell)})\,,
\end{align}
which allows one to compute the network output recursively.
Both the parameters $\phi$ and $\theta$ will be learned to solve the following optimization problem:
\begin{align}
    \label{eq:objective}
    \min_{\theta,\phi}
    {\cal R}(\theta)
    +
    {\cal R}'(\phi)
    -
    {\cal L}
    \,.
\end{align}
where ${\cal R}(\theta), {\cal R}'(\phi)$
are regularization terms for the parameters $\theta, \phi$.
We call this model a generalized probabilistic binary neural network, with $\phi$ deformation parameters chosen such that $\phi=0$ gives back the standard probabilistic binary neural network.

To study this model on a classical computer we need to choose $S$ which leads to an efficient sampling algorithm for $\pi_{\phi,\theta}$. 
In general, one could use Markov Chain Monte Carlo, but there exists situations for which the mixing time
of the chain grows exponentially in the size of the problem \citep{levin2017markov}. In the next section we will show how quantum mechanics can enlarge the set of probabilistic binary neural networks that can be efficiently executed and in the subsequent sections we will show experimental results for a restricted class of correlated distributions inspired by quantum circuits that can be simulated classically.

\section{Quantum implementation}

Quantum computers can sample from certain correlated distributions more efficiently than classical computers
\citep{aaronson2016complexitytheoretic,arute2019quantum}.
In this section, we devise a quantum circuit that implements the generalized probabilistic binary neural networks introduced above, encoding $\pi_{\theta,\phi}$ in a quantum circuit. This leads to an exponential speedup for running this model on a quantum computer, opening up the study of more complex probabilistic neural networks.

A quantum implementation of a binary perceptron was introduced in \cite{Schuld_2015} as an application of the quantum phase estimation algorithm \citep{NielsenChuang}.
However, no quantum advantage of the quantum simulation was shown.
Here we will extend that result in several ways: i) we will modify the algorithm to represent the generalized probabilistic layer introduced above, showing the quantum advantage present in our setting; ii) we will consider the case of multi layer percetrons as well as convolutional networks.

\subsection{Introduction to quantum notation and quantum phase estimation}
\label{sec:qpe}

As a preliminary step, we introduce notations for quantum mechanics.
We refer the reader to Appendix \ref{sec:qm} for a more thorough review of quantum mechanics.
A qubit is the vector space of normalized vectors $\ket{\psi} \in \mathbb{C}^2$. 
$N$ qubits form the set of unit vectors in 
$(\mathbb{C}^2)^{\otimes N} \cong \mathbb{C}^{2^N}$ spanned by all $N$-bit strings, $\ket{b_1, \dots, b_N}\equiv \ket{b_1} \otimes \cdots \otimes \ket{b_N}$, $b_i\in \mathbb{B}$. Quantum circuits are unitary matrices on this space. 
The probability of a measurement with outcome $\phi_i$ is given by matrix element of the projector $\ket{\phi_i}\bra{\phi_i}$ in a state $\ket{\psi}$, namely $p_i=\bra{\psi}\ket{\phi_i}\bra{\phi_i}\ket{\psi}
=|\bra{\phi_i}\ket{\psi}|^2$, a formula known as Born's rule. 

Next, we describe the quantum phase estimation (QPE), a quantum algorithm to estimate the eigenphases of a unitary $\mU$. 
Denote the eigenvalues and eigenvectors of $\mU$
by $\exp(\tfrac{2\pi i}{2^t}\varphi_\alpha)$ and $\ket{v_\alpha}$, and assume that the $\varphi_\alpha$'s can be represented with a finite number  
$t$ of bits: $\varphi_\alpha = 2^{t-1} \varphi_\alpha^1 + \dots + 2^{0} \varphi_\alpha^t$.
(This is the case of relevance for a binary network.)
Then introduce $t$ ancilla qubits in state $\ket{0}^{\otimes t}$. Given an input state $\ket{\psi}$, QPE is the following unitary operation:
\begin{align}
    \label{eq:qpe_psi}
    \ket{0}^{\otimes t}
    \otimes 
    \ket{\psi}
    \overset{\text{QPE}}{\longmapsto}
    \sum_{\alpha}
    \braket{v_\alpha}{\psi}
    \ket{\varphi_\alpha}\otimes \ket{v_\alpha} \,.
\end{align}
Appendix \ref{sec:qpe_details} reviews the details of the quantum circuit implementing this map, whose
complexity is linear in $t$. 
Now using the notation $\tau$ for the threshold non-linearity introduced in \eqref{eq:hell}, and recalling the expansion $2^{-t}\varphi = 
2^{-1} \varphi^1 + \dots + 2^{-t} \varphi^t$, we note that if the first bit $\varphi^1 = 0$ then $2^{-t}\varphi < \frac{1}{2}$ and $\tau(2^{-t}\varphi)=0$, while if 
$\varphi^1 = 1$, then 
$2^{-t}\varphi \ge \frac{1}{2}$ and $\tau(2^{-t}\varphi)=1$.
In other words, $\delta_{\varphi^1,b}=\delta_{\tau(2^{-t}\varphi),b}$ and
the probability $p(b)$ that after the QPE the first ancilla bit is $b$ is given by:
\begin{align}
\label{eq:p_first_bit}
\Big(\!\sum_{\alpha} \overline{\braket{v_\alpha}{\psi}}
\bra{\varphi_\alpha}\otimes \bra{v_\alpha}\!\Big)
\!\!\Big[\ket{b}\!\bra{b}\!\otimes\! {\bf 1}\Big]
\!\Big(\!\sum_{\beta}
\braket{v_\beta}{\psi}
\ket{\varphi_\beta}\!\otimes \!\ket{v_\beta} \!\Big)
=
\sum_\alpha |\!\braket{v_\alpha}{\psi}\!|^2 \delta_{\tau(2^{-t}\varphi_\alpha),b}
\,,
\end{align}
where $\Big[\ket{b}\!\bra{b}\!\otimes\! {\bf 1}\Big]$ is an operator that projects the first bit to the state $\ket{b}$ and leaves the other bits untouched.

\begin{figure}[ht]
\begin{center}
\vspace{-1cm}
\mbox{{\scriptsize 
\providecommand{\mynus}[1][.7]{\scalebox{#1}{-}}
\begin{tikzpicture}[scale=0.850000,x=1pt,y=1pt]
\filldraw[color=white] (0.000000, -7.500000) rectangle (95.000000, 97.500000);
\draw[color=black] (0.000000,90.000000) -- (95.000000,90.000000);
\draw[color=black] (0.000000,90.000000) node[left] {$\ket{0}^{\otimes t}$};
\draw[color=black] (0.000000,75.000000) node[anchor=mid east] {$\vdots$};
\draw[color=black] (0.000000,60.000000) -- (95.000000,60.000000);
\draw[color=black] (0.000000,60.000000) node[left] {$\ket{0}^{\otimes t}$};
\draw[color=black] (0.000000,45.000000) -- (95.000000,45.000000);
\draw[color=black] (0.000000,45.000000) node[left] {$\ket{\psi_h}$};
\draw[color=black] (0.000000,30.000000) -- (95.000000,30.000000);
\draw[color=black] (0.000000,30.000000) node[left] {$|\psi_{\mW_{1,:}}\rangle$};
\draw[color=black] (0.000000,15.000000) node[anchor=mid east] {$\vdots$};
\draw[color=black] (0.000000,0.000000) -- (95.000000,0.000000);
\draw[color=black] (0.000000,0.000000) node[left] {$|\psi_{\mW_{M,:}}\rangle$};
\draw (32.000000,60.000000) -- (32.000000,0.000000);
\begin{scope}
\draw[fill=white] (32.000000, 30.000000) +(-45.000000:8.485281pt and 50.911688pt) -- +(45.000000:8.485281pt and 50.911688pt) -- +(135.000000:8.485281pt and 50.911688pt) -- +(225.000000:8.485281pt and 50.911688pt) -- cycle;
\clip (32.000000, 30.000000) +(-45.000000:8.485281pt and 50.911688pt) -- +(45.000000:8.485281pt and 50.911688pt) -- +(135.000000:8.485281pt and 50.911688pt) -- +(225.000000:8.485281pt and 50.911688pt) -- cycle;
\draw (32.000000, 30.000000) node {\rotatebox{90}{QPE$(\mU_1)$}};
\end{scope}
\draw[color=black] (57.500000, 90.000000) node [fill=white] {$\cdots$};
\draw[color=black] (57.500000, 60.000000) node [fill=white] {$\cdots$};
\draw[color=black] (57.500000, 45.000000) node [fill=white] {$\cdots$};
\draw[color=black] (57.500000, 30.000000) node [fill=white] {$\cdots$};
\draw[color=black] (57.500000, 0.000000) node [fill=white] {$\cdots$};
\draw (83.000000,90.000000) -- (83.000000,0.000000);
\begin{scope}
\draw[fill=white] (83.000000, 90.000000) +(-45.000000:8.485281pt and 8.485281pt) -- +(45.000000:8.485281pt and 8.485281pt) -- +(135.000000:8.485281pt and 8.485281pt) -- +(225.000000:8.485281pt and 8.485281pt) -- cycle;
\clip (83.000000, 90.000000) +(-45.000000:8.485281pt and 8.485281pt) -- +(45.000000:8.485281pt and 8.485281pt) -- +(135.000000:8.485281pt and 8.485281pt) -- +(225.000000:8.485281pt and 8.485281pt) -- cycle;
\draw (83.000000, 90.000000) node {${}$};
\end{scope}
\begin{scope}
\draw[fill=white] (83.000000, 22.500000) +(-45.000000:8.485281pt and 40.305087pt) -- +(45.000000:8.485281pt and 40.305087pt) -- +(135.000000:8.485281pt and 40.305087pt) -- +(225.000000:8.485281pt and 40.305087pt) -- cycle;
\clip (83.000000, 22.500000) +(-45.000000:8.485281pt and 40.305087pt) -- +(45.000000:8.485281pt and 40.305087pt) -- +(135.000000:8.485281pt and 40.305087pt) -- +(225.000000:8.485281pt and 40.305087pt) -- cycle;
\draw (83.000000, 22.500000) node {\rotatebox{90}{QPE$(\mU_M)$}};
\end{scope}
\draw[color=black] (95.000000,75.000000) node[anchor=mid west] {$\vdots$};
\draw[color=black] (95.000000,15.000000) node[anchor=mid west] {$\vdots$};
\draw[color=black] (32.000000,-15.000000) node[anchor=mid west] {(a)};
\end{tikzpicture}
\providecommand{\mynus}[1][.7]{\scalebox{#1}{-}}
\begin{tikzpicture}[scale=.9500000,x=1pt,y=1pt]
\filldraw[color=white] (0.000000, -7.500000) rectangle (96.000000, 142.500000);
\draw[color=black] (0.000000,135.000000) -- (84.000000,135.000000);
\draw[color=black] (84.000000,134.500000) -- (96.000000,134.500000);
\draw[color=black] (84.000000,135.500000) -- (96.000000,135.500000);
\draw[color=black] (0.000000,135.000000) node[left] {$\ket{0}$};
\draw[color=black] (0.000000,120.000000) -- (96.000000,120.000000);
\draw[color=black] (0.000000,120.000000) node[left] {$\ket{0}^{\otimes t_2\mynus1}$};
\draw[color=black] (0.000000,105.000000) -- (96.000000,105.000000);
\draw[color=black] (0.000000,105.000000) node[left] {$\ket{0}$};
\draw[color=black] (0.000000,90.000000) -- (96.000000,90.000000);
\draw[color=black] (0.000000,90.000000) node[left] {$\ket{0}^{\otimes t_1\mynus1}$};
\draw[color=black] (0.000000,75.000000) -- (96.000000,75.000000);
\draw[color=black] (0.000000,75.000000) node[left] {$\ket{0}$};
\draw[color=black] (0.000000,60.000000) -- (96.000000,60.000000);
\draw[color=black] (0.000000,60.000000) node[left] {$\ket{0}^{\otimes t_1\mynus1}$};
\draw[color=black] (0.000000,45.000000) -- (96.000000,45.000000);
\draw[color=black] (0.000000,45.000000) node[left] {$\ket{\bm{x}}$};
\draw[color=black] (0.000000,30.000000) -- (96.000000,30.000000);
\draw[color=black] (0.000000,30.000000) node[left]
{$|\psi_{\mW^1_{1,:}}\rangle$};
\draw[color=black] (0.000000,15.000000) -- (96.000000,15.000000);
\draw[color=black] (0.000000,15.000000) node[left] {$|\psi_{\mW^1_{2,:}}\rangle$};
\draw[color=black] (0.000000,0.000000) -- (96.000000,0.000000);
\draw[color=black] (0.000000,0.000000) node[left] {$|\psi_{\mW^2_{1,:}}\rangle$};
\draw (12.000000,75.000000) -- (12.000000,15.000000);
\begin{scope}
\draw[fill=white] (12.000000, 45.000000) +(-45.000000:8.485281pt and 50.911688pt) -- +(45.000000:8.485281pt and 50.911688pt) -- +(135.000000:8.485281pt and 50.911688pt) -- +(225.000000:8.485281pt and 50.911688pt) -- cycle;
\clip (12.000000, 45.000000) +(-45.000000:8.485281pt and 50.911688pt) -- +(45.000000:8.485281pt and 50.911688pt) -- +(135.000000:8.485281pt and 50.911688pt) -- +(225.000000:8.485281pt and 50.911688pt) -- cycle;
\draw (12.000000, 45.000000) node {\rotatebox{90}{QPE$(\mU^1_1)$}};
\end{scope}
\draw (36.000000,105.000000) -- (36.000000,15.000000);
\begin{scope}
\draw[fill=white] (36.000000, 97.500000) +(-45.000000:8.485281pt and 19.091883pt) -- +(45.000000:8.485281pt and 19.091883pt) -- +(135.000000:8.485281pt and 19.091883pt) -- +(225.000000:8.485281pt and 19.091883pt) -- cycle;
\clip (36.000000, 97.500000) +(-45.000000:8.485281pt and 19.091883pt) -- +(45.000000:8.485281pt and 19.091883pt) -- +(135.000000:8.485281pt and 19.091883pt) -- +(225.000000:8.485281pt and 19.091883pt) -- cycle;
\draw (36.000000, 97.500000) node {${}$};
\end{scope}
\begin{scope}
\draw[fill=white] (36.000000, 30.000000) +(-45.000000:8.485281pt and 29.698485pt) -- +(45.000000:8.485281pt and 29.698485pt) -- +(135.000000:8.485281pt and 29.698485pt) -- +(225.000000:8.485281pt and 29.698485pt) -- cycle;
\clip (36.000000, 30.000000) +(-45.000000:8.485281pt and 29.698485pt) -- +(45.000000:8.485281pt and 29.698485pt) -- +(135.000000:8.485281pt and 29.698485pt) -- +(225.000000:8.485281pt and 29.698485pt) -- cycle;
\draw (36.000000, 30.000000) node {\rotatebox{90}{QPE$(\mU^1_2)$}};
\end{scope}
\draw (60.000000,135.000000) -- (60.000000,0.000000);
\begin{scope}
\draw[fill=white] (60.000000, 120.000000) +(-45.000000:8.485281pt and 29.698485pt) -- +(45.000000:8.485281pt and 29.698485pt) -- +(135.000000:8.485281pt and 29.698485pt) -- +(225.000000:8.485281pt and 29.698485pt) -- cycle;
\clip (60.000000, 120.000000) +(-45.000000:8.485281pt and 29.698485pt) -- +(45.000000:8.485281pt and 29.698485pt) -- +(135.000000:8.485281pt and 29.698485pt) -- +(225.000000:8.485281pt and 29.698485pt) -- cycle;
\draw (60.000000, 120.000000) node {\rotatebox{90}{QPE$(\mU^2_1)$}};
\end{scope}
\begin{scope}
\draw[fill=white] (60.000000, 75.000000) +(-45.000000:8.485281pt and 8.485281pt) -- +(45.000000:8.485281pt and 8.485281pt) -- +(135.000000:8.485281pt and 8.485281pt) -- +(225.000000:8.485281pt and 8.485281pt) -- cycle;
\clip (60.000000, 75.000000) +(-45.000000:8.485281pt and 8.485281pt) -- +(45.000000:8.485281pt and 8.485281pt) -- +(135.000000:8.485281pt and 8.485281pt) -- +(225.000000:8.485281pt and 8.485281pt) -- cycle;
\draw (60.000000, 75.000000) node {${}$};
\end{scope}
\begin{scope}
\draw[fill=white] (60.000000, -0.000000) +(-45.000000:8.485281pt and 8.485281pt) -- +(45.000000:8.485281pt and 8.485281pt) -- +(135.000000:8.485281pt and 8.485281pt) -- +(225.000000:8.485281pt and 8.485281pt) -- cycle;
\clip (60.000000, -0.000000) +(-45.000000:8.485281pt and 8.485281pt) -- +(45.000000:8.485281pt and 8.485281pt) -- +(135.000000:8.485281pt and 8.485281pt) -- +(225.000000:8.485281pt and 8.485281pt) -- cycle;
\draw (60.000000, -0.000000) node {${}$};
\end{scope}
\draw[fill=white] (78.000000, 129.000000) rectangle (90.000000, 141.000000);
\draw[very thin] (84.000000, 135.600000) arc (90:150:6.000000pt);
\draw[very thin] (84.000000, 135.600000) arc (90:30:6.000000pt);
\draw[->,>=stealth] (84.000000, 129.600000) -- +(80:10.392305pt);
\draw[color=black] (96.000000,135.000000) node[right] {$y$};
\draw[dashed] (48.000000, -7.500000) -- (48.000000, 142.500000);
\draw[dashed] (72.000000, -7.500000) -- (72.000000, 142.500000);
\draw[decorate,decoration={brace,mirror,amplitude = 4.000000pt},very thick] (3.000000,-7.500000) -- (45.000000,-7.500000);
\draw (24.000000, -11.500000) node[text width=144pt,below,text centered] {Layer 1};
\draw[decorate,decoration={brace,mirror,amplitude = 4.000000pt},very thick] (51.000000,-7.500000) -- (69.000000,-7.500000);
\draw (60.000000, -11.500000) node[text width=144pt,below,text centered] {Layer 2};
\draw[color=black] (40.000000,-25.000000) node[anchor=mid west] {(b)};
\end{tikzpicture}
\hspace{-1cm}
\providecommand{\mynus}[1][.7]{\scalebox{#1}{-}}
\begin{tikzpicture}[scale=.850000,x=1pt,y=1pt]
\filldraw[color=white] (0.000000, -7.500000) rectangle (44.000000, 97.500000);
\draw[color=black] (0.000000,90.000000) -- (44.000000,90.000000);
\draw[color=black] (0.000000,90.000000) node[left] {$\ket{0}^{\otimes t}$};
\draw[color=black] (0.000000,75.000000) -- (44.000000,75.000000);
\draw[color=black] (0.000000,75.000000) node[left] {$\ket{\psi_h}$};
\draw[color=black] (0.000000,60.000000) -- (44.000000,60.000000);
\draw[color=black] (0.000000,60.000000) node[left] {$|\psi_{\mW_{1,:}}\rangle$};
\draw[color=black] (0.000000,45.000000) node[anchor=mid east] {$\vdots$};
\draw[color=black] (0.000000,30.000000) -- (44.000000,30.000000);
\draw[color=black] (0.000000,30.000000) node[left] {$\ket{0}^{\otimes t}$};
\draw[color=black] (0.000000,15.000000) -- (44.000000,15.000000);
\draw[color=black] (0.000000,15.000000) node[left] {$\ket{\psi_h}$};
\draw[color=black] (0.000000,0.000000) -- (44.000000,0.000000);
\draw[color=black] (0.000000,0.000000) node[left] {$|\psi_{\mW_{M,:}}\rangle$};
\draw (32.000000,90.000000) -- (32.000000,60.000000);
\begin{scope}
\draw[fill=white] (32.000000, 75.000000) +(-45.000000:8.485281pt and 29.698485pt) -- +(45.000000:8.485281pt and 29.698485pt) -- +(135.000000:8.485281pt and 29.698485pt) -- +(225.000000:8.485281pt and 29.698485pt) -- cycle;
\clip (32.000000, 75.000000) +(-45.000000:8.485281pt and 29.698485pt) -- +(45.000000:8.485281pt and 29.698485pt) -- +(135.000000:8.485281pt and 29.698485pt) -- +(225.000000:8.485281pt and 29.698485pt) -- cycle;
\draw (32.000000, 75.000000) node {\rotatebox{90}{QPE$(\tilde{\mU}_1)$}};
\end{scope}
\draw (32.000000,30.000000) -- (32.000000,0.000000);
\begin{scope}
\draw[fill=white] (32.000000, 15.000000) +(-45.000000:8.485281pt and 29.698485pt) -- +(45.000000:8.485281pt and 29.698485pt) -- +(135.000000:8.485281pt and 29.698485pt) -- +(225.000000:8.485281pt and 29.698485pt) -- cycle;
\clip (32.000000, 15.000000) +(-45.000000:8.485281pt and 29.698485pt) -- +(45.000000:8.485281pt and 29.698485pt) -- +(135.000000:8.485281pt and 29.698485pt) -- +(225.000000:8.485281pt and 29.698485pt) -- cycle;
\draw (32.000000, 15.000000) node {\rotatebox{90}{QPE$(\tilde{\mU}_M)$}};
\end{scope}
\draw[color=black] (44.000000,45.000000) node[anchor=mid west] {$\vdots$};
\draw[color=black] (5.000000,-15.000000) node[anchor=mid west] {(c)};
\end{tikzpicture}
}}
\end{center}
\vspace{-0.5cm}
\caption{(a) Quantum circuit implementing a quantum deformed layer. The thin vertical line indicates that the gate acts as identity on the wires crossed by the line.
(b) Quantum deformed multilayer perceptron with $2$ hidden quantum neurons and $1$ output quantum neuron. $\ket{\bm{x}}$ is an encoding of the input signal, $y$ is the prediction. The superscript $\ell$ in $\mU_j^\ell$ and $\mW_{j,:}^\ell$ refers to layer $\ell$.
We split the blocks of $t_\ell$ ancilla qubits into a readout qubit that encodes the layer output amplitude and the rest.
(c) Modification of a layer for classical simulations.}
\label{fig:circuits}
\end{figure}
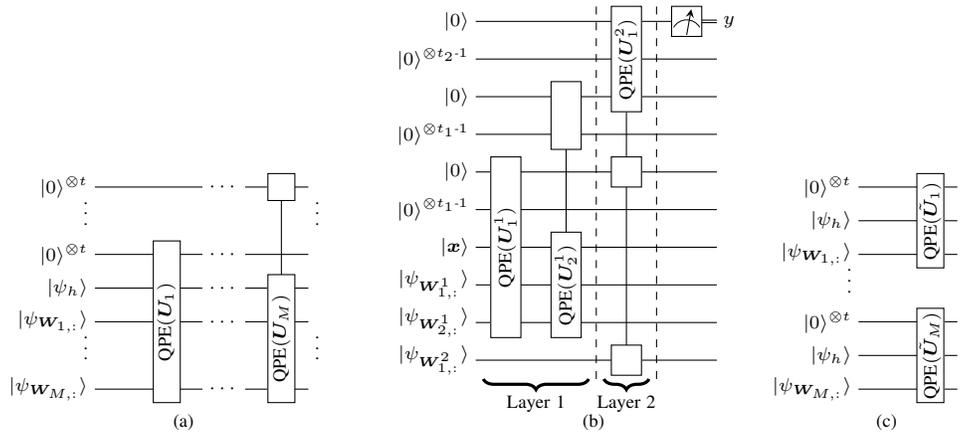

\subsection{Definition and advantages of quantum deformed neural networks}
\label{sec:qdnn}

Armed with this background, we can now apply quantum phase estimation to compute the output of the probabilistic layer of \eqref{eq:py_S}. Let $N$ be the number of input neurons and $M$ that of output neurons. We introduce qubits to represent inputs and weights bits:
\begin{align}
\ket{\bm{h}, \mW} \in 
{\cal V}_h
\otimes
{\cal V}_W
\,,\quad
{\cal V}_h = \bigotimes_{i=1}^N (\mathbb{C}^2)_i
\,,\quad
{\cal V}_W =
\bigotimes_{i=1}^N 
\bigotimes_{j=1}^M (\mathbb{C}^2)_{ij}
\,.
\end{align}
Then we introduce a Hamiltonian $\mH_j$ acting non trivially only on the $N$ input activations and the $N$ weights at the $j$-th row:
\begin{align}
\label{eq:H}
\mH_j = \sum_{i=1}^N \mB_{ji}^W \mB_{i}^h\,,
\end{align}
and $\mB_i^h$ ($\mB_{ji}^W$) is the matrix $\mB = \ket{1}\bra{1}$ acting on the $i$-th activation ($ji$-th weight) qubit.
Note that $\mH_j$ singles out terms from the state $\ket{\bm{h},\mW}$ where both $h_j=1$ and $W_{ij}=1$ and then adds them up, i.e. the eigenvalues of $\mH_j$ are the preactivations of \eqref{eq:hell}:
\begin{align}
\label{eq:varphi}
\mH_j \ket{\bm{h}, \mW} = 
\varphi(\bm{h},\bm{W}_{j,:})
\ket{\bm{h}, \mW}
\,,
\quad
\varphi(\bm{h},\bm{W}_{j,:})
=
\sum_{i=1}^N W_{ji}h_i
\,.
\end{align}
Now define the unitary operators:
\begin{align}
\label{eq:u_j}
\mU_j = \mD {\rm e}^{\frac{2\pi i}{N+1} \mH_j} \mD^{-1}\,,
\end{align}
where $\mD$ is another generic unitary, and as we shall see shortly, its eigenvectors will be related to the entries of the classical stochastic matrix $\mS$ in section \ref{sec:generalization}.
Since $\mU_j\mU_{j'} = 
\mD {\rm e}^{\frac{2\pi i}{N+1} (\mH_{j}+\mH_{j'})} \mD^{-1}
=
\mU_{j'}\mU_j$, we can diagonalize all the $\mU_j$'s simultaneously and since they are conjugate to ${\rm e}^{\frac{2\pi i}{N+1} \mH_j}$ they will have the same eigenvalues. Introducing the eigenbasis 
$\ket{\bm{h},\bm{W}}_{\mD}=\mD\ket{\bm{h},\bm{W}}$, we have:
\begin{align}
    \mU_j 
    \ket{\bm{h},\bm{W}}_{\mD} 
    &= 
    {\rm e}^{\frac{2\pi i}{N+1} \varphi(\bm{h},\bm{W}_{j,:})}
    \ket{\bm{h},\bm{W}}_{\mD}
    \,.
\end{align}
Note that $\varphi\in \{0,\dots,N\}$ so we can represent it with exactly $t$ bits, $N=2^t - 1$.
Then we add $M$ ancilla resources, each of $t$ qubits, and sequentially perform $M$ quantum phase estimations, one for each $\mU_j$, as
depicted in \figref{fig:circuits} (a).
We choose the following input state 
\begin{align}
    \label{eq:psi}
    \ket{\psi}
    =
    \ket{\psi}_h
    \otimes 
    \bigotimes_{j=1}^M
    \ket{\psi}_{\mW_{j,:}}
    \,,\quad
    \ket{\psi}_{\mW_{j,:}}
    =
    \bigotimes_{i=1}^N
    \left[\sqrt{q_{ji}(W_{ji}=0)} \ket{0} + \sqrt{q_{ji}(W_{ji}=1)} \ket{1}\right]
    \,,
\end{align}
where we have chosen the weight input state according to the factorized variational distribution $q_{ij}$ introduced in section \ref{sec:generalization}.
In fact, this state corresponds to the following probability distribution via Born's rule:
\begin{align}
    p(\bm{h}, \mW)
    =
    |\braket{\bm{h}, \mW}{\psi}|^2
    =
    p(\bm{h})
    \prod_{j=1}^M \prod_{i=1}^N
    q_{ji}(W_{ji})
    \,,\quad
    p(\bm{h})
    =
    |\braket{\bm{h}}{\psi}_h|^2\,.
\end{align}
The state $\ket{\psi_h}$ is discussed below.
Now we show that a non-trivial choice of $\mD$ leads to an effective correlated distribution. 
The $j$-th QPE in \figref{fig:circuits} (a) corresponds to \eqref{eq:qpe_psi} where we identify
$\ket{v_\alpha} \equiv \ket{\bm{h},\mW}_{\mD}$, $\ket{\varphi_\alpha}\equiv\ket{\varphi(\bm{h},\mW_{j,:})}$ and we make use of the $j$-th block of $t$ ancillas.
After $M$ steps we 
compute the outcome probability of a measurement of the first qubit in each of the $M$ registers of the ancillas. 
We can extend \eqref{eq:p_first_bit} to the situation of measuring multiple qubits, 
and recalling that the first bit of an integer is the most significant bit, determining whether $2^{-t} \varphi(\bm{h},\mW_{j,:}) = (N+1)^{-1} \varphi(\bm{h},\mW_{j,:})$ is greater or smaller than $1/2$, the probability of outcome $\bm{h}'=(h'_1,\dots,h_M')$ is
\begin{align}
    \label{eq:py_D}
    p(\bm{h}') 
    &= 
    \sum_{\bm{h}\in \mathbb{B}^N}
    \sum_{\mW\in \mathbb{B}^{N M}}
    \delta_{\bm{h}' , f(\mW, \bm{h})}
    \left|\braket{\psi}{\bm{h},\mW}_{\mD}\right|^2\,,
\end{align}
where $f$ is the layer function introduced in \eqref{eq:hell}.
We refer to appendix \ref{sec:details_qdnn} for a detailed derivation.
Equation \ref{eq:py_D} is the generalized probabilistic binary layer introduced in \eqref{eq:py_S} where $\mD$ corresponds to a non-trivial $\mS$ and a correlated distribution when $\mD$ entangles the qubits:
\begin{align}
\label{eq:pi_D}
\pi(\bm{h},\mW) 
= 
\left|\bra{\psi}\mD\ket{\bm{h},\mW}\right|^2
\,.
\end{align}
The variational parameters $\phi$ of $\mS$ are now parameters of the quantum circuit $\mD$.
Sampling from $\pi$ can be done by doing repeated measurements of the first $M$ ancilla qubits of this quantum circuit. On quantum hardware ${\rm e}^{\frac{2\pi i}{N+1} \mH_j}$ can be efficiently implemented since it is a product of diagonal two-qubits quantum gates. We shall consider unitaries $\mD$ which have efficient quantum circuit approximations. Then computing the quantum deformed layer output on a quantum computer is going to take time $O(t M u(N))$ where $u(N)$ is the time it takes to compute the action of $\mU_j$ on an input state.
There exists $\mD$ such that sampling from \eqref{eq:pi_D} is exponentially harder classically than quantum mechanically, a statement forming the basis for quantum supremacy experiments on noisy, intermediate scale quantum computers \citep{aaronson2016complexitytheoretic,arute2019quantum}.
Examples are random circuits with two-dimensional entanglement patterns, which from a machine learning point of view can be natural when considering image data.
Other examples are $\mD$ implementing time evolution operators of physical systems, whose simulation is exponentially hard classically, resulting in hardness of sampling from the time evolved wave function.
Quantum supremacy experiments give foundations to which architectures can benefit from quantum speedups, but we remark that the proposed quantum  architecture, which relies on quantum phase estimation, is designed for error-corrected quantum computers.

Even better, on quantum hardware we can avoid sampling intermediate activations altogether.
At the first layer, the input can be prepared by encoding the input bits in the state $\ket{\bm{x}}$.
For the next layers, we simply use the output state as the input to the next layer. One obtains thus the quantum network of \figref{fig:circuits} (b) and the algorithm for a layer is summarized in procedure \ref{algq}.
Note that all the qubits associated to the intermediate activations are entangled. Therefore the input state $\ket{\psi_h}$ would have to be replaced by a state in ${\cal V}_h$ plus all the other qubits, where the gates at the next layer would act only on ${\cal V}_h$ in the manner described in this section. (An equivalent and more economical mathematical description is to use the reduced density matrix $\rho_h$ as input state.)
We envision two other possible procedures 
for what happens after the first layer: i) we sample from \eqref{eq:py_D} and initialize $\ket{\psi_h}$ to the bit string sampled in analogy to the classical quantization of activations; ii) we sample many times to reconstruct the classical distribution and encode it in $\ket{\psi_h}$.
In our classical simulations below we will be able to actually calculate the probabilities and can avoid sampling. 

Finally, we remark that at present it is not clear whether the computational speedup exhibited by our architecture translates to a learning advantage.
This is an outstanding question whose full answer will require an empirical evaluation with a quantum computer. 
Next, we will try to get as close as possible to answer this question by studying a quantum model that we can simulate classically.

\subsection{Modifications for classical simulations}
\label{sec:modifs_cl}
In this paper we will provide classical simulations of the quantum neural networks introduced above for a restricted class of designs. We do this for two reasons: first to convince the reader that the quantum layers hold promise (even though we can not simulate the proposed architecture in its full glory due to the lack of access to a quantum computer) and second, to show that these ideas can be interesting as new designs, even ``classically'' (by which we mean architectures that can be executed on a classical computer). 

To parallelize the computations for different output neurons, we do the modifications to the setup just explained which are depicted in \figref{fig:circuits} (c). We clone the input activation register $M$ times, an operation that quantum mechanically is only approximate \citep{NielsenChuang} but exact classically.
Then we associate the $j$-th copy to the $j$-th row of the weight matrix, thus forming pairs 
for each $j=1,\dots, M$:
\begin{align}
\ket{\bm{h}, \mW_{j,:}} \in 
{\cal V}_h\otimes {\cal V}_{W,j}
\,,\quad
{\cal V}_{W,j} = \bigotimes_{i=1}^N (\mathbb{C}^2)_{ji}
\end{align}
Fixing $j$, we introduce the unitary ${\rm e}^{\frac{2\pi i}{N+1} \mH_j}$ diagonal in 
the basis $\ket{\bm{h}, \mW_{j,:}}$ as in \eqref{eq:H} and define the new unitary:
\begin{align}
\label{eq:u_tilde}
\tilde{\mU}_j = \mD_j {\rm e}^{\frac{2\pi i}{N+1} \mH_j} \mD_j^{-1}\,,
\end{align}
where w.r.t.~\eqref{eq:u_j} we now let $\mD_j$ depend on $j$.
We denote the eigenvectors of $\tilde{\mU}_j$ by 
$\ket{\bm{h},\bm{W}_{j,:}}_{\mD_j} = \mD_j 
\ket{\bm{h},\bm{W}_{j,:}}$ and the eigenvalue is 
$\varphi(\bm{h},\mW_{j,:})$ introduced in \eqref{eq:varphi}.
Supposing that we know $p(\bm{h}) =\prod_i p_i(h_i)$, we apply the quantum phase estimation to $\tilde{\mU}_j$ with input:
\begin{align}
    \label{eq:psi_j}
    \ket{\psi_j}
    =
    \ket{\psi_h}
    \otimes
    \ket{\psi}_{\mW_{j,:}}
    \,,
    \quad
    \ket{\psi_h}
    =
    \bigotimes_{i=1}^N
    \left[\sqrt{p_i(h_i=0)} \ket{0} + \sqrt{p_i(h_i=1)} \ket{1}\right]
    \,,
\end{align}
and $\ket{\psi}_{\mW_{j,:}}$ is defined in \eqref{eq:psi}.
Going through similar calculations as those done above shows that measurements of the first qubit will be governed by the probability distribution of \eqref{eq:py_S} factorized over output channels since the procedure does not couple them:
$\pi(\bm{h},\mW) 
= 
\prod_{j=1}^M
\left| 
\bra{\psi_j} \mD_j \ket{\bm{h},\mW_{j,:}}
\right|^2$.
So far, we have focused on fully connected layers.
We can extend the derivation of this section to the convolution case, by applying the quantum phase estimation on images patches of the size equal to the kernel size as explained in appendix \ref{sec:case_conv}.

\section{Classical simulations for low entanglement}

\subsection{Theory}

It has been remarked in \citep{shayer,pbnn} that when 
the weight and activation distributions at a given layer are factorized, 
$p(\bm{h}) = \prod_i p_i(h_i)$ and
$q(\bm{W}) = \prod_{ij} q_{ij}(W_{ij})$,
the output distribution in \eqref{eq:py} can be efficiently approximated using the central limit theorem (CLT). The argument goes as follows: for each $j$ the preactivations $\varphi(\bm{h},\mW_{j,:}) = \sum_{i=1}^N W_{j, i} h_i$ are sums of independent binary random variables $W_{j, i} h_i$ with mean and variance:
\begin{align}
    \mu_{ji} =
    \mathbb{E}_{w\sim q_{ji}}(w)
    \mathbb{E}_{h\sim p_{i}}(h)\,,
    \quad
    \sigma^2_{ji} =
    \mathbb{E}_{w\sim q_{ji}}(w^2)
    \mathbb{E}_{h\sim p_{i}}(h^2)
    -
    \mu_{ji}^2
    =
    \mu_{ji}(1-\mu_{ji})
    \,,
\end{align}
We used $b^2=b$ for a variable $b\in\{0,1\}$.
The CLT implies that for large $N$ we can approximate 
$\varphi(\bm{h},\mW_{j,:})$ with a normal distribution with mean 
$\mu_j = \sum_{i}\mu_{ji}$ and variance $\sigma_j^2 = \sum_{i}\sigma_{ji}^2$. The distribution of the activation after the non-linearity of \eqref{eq:hell} can thus be computed as:
\begin{align}
    \label{eq:p_cdf}
    p(\tau(\tfrac{1}{N+1}
    \varphi(\bm{h},\mW_{j,:})) = 1)
    =
    p(2\varphi(\bm{h},\mW_{j,:})-N > 0)
    =
    \Phi\left(- \tfrac{2\mu_j-N}{2\sigma_j} \right)
    \,,
\end{align}
$\Phi$ being the CDF of the standard normal distribution. Below we fix $j$ and omit it for notation clarity.

As reviewed in appendix \ref{sec:qm}, commuting observables in quantum mechanics behave like classical random variables. The observable of interest for us, $\mD \mH \mD^{-1}$ of \eqref{eq:u_tilde}, is a sum of commuting terms $\mK_i \equiv \mD \mB_{i}^W \mB_{i}^h \mD^{-1}$ and if their joint probability distribution is such that
these random variables are weakly correlated, i.e.
\begin{align}
    \bra{\psi} \mK_i \mK_{i'} \ket{\psi}
    -
    \bra{\psi} \mK_i \ket{\psi}\bra{\psi} \mK_{i'} \ket{\psi}
    \to 0 \,,\quad \text{if } |i-i'| \to \infty\,,
\end{align}
then the CLT for weakly correlated random variables applies, stating that measurements of $\mD \mH \mD^{-1}$ in state $\ket{\psi}$ are governed by a Gaussian distribution ${\cal N}(\mu, \sigma^2)$ with
\begin{align}
    \label{eq:mu_sigma}
    \mu = \bra{\psi} \mD \mH \mD^{-1} \ket{\psi}
    \,,\quad
    \sigma^2 = 
    \bra{\psi} \mD \mH^2 \mD^{-1} \ket{\psi}
    -
    \mu^2
    \,.
\end{align}
Finally, we can plug these values into \eqref{eq:p_cdf} to get the layer output probability.

We have cast the problem of simulating the quantum neural network to the problem of computing the expectation values in \eqref{eq:mu_sigma}. In physical terms, these are related to correlation functions of $\mH$ and $\mH^2$ after evolving a state $\ket{\psi}$ with the operator $\mD$. These can be efficiently computed classically for one dimensional and lowly entangled quantum circuits $\mD$ \citep{vidal2003efficient}. In view of that here we consider a 1d arrangement of activation and weight qubits, labeled by $i=0,\dots,2N-1$, where the even qubits are associated with activations and the odd are associated with weights. We then choose:
\begin{align}
    \label{eq:D_Q_P}
    \mD
    =
    \prod_{i=0}^{N-1}
    \mQ_{2i, 2i+1}
    \prod_{i=0}^{N-1}
    \mP_{2i+1, 2i+2}
    \,,
\end{align}
where $\mQ_{2i, 2i+1}$ acts non-trivially on qubits $2i,2i+1$, i.e. onto the $i$-th activation and $i$-th weight qubits, while 
$\mP_{2i, 2i+1}$ on the $i$-th weight and $i+1$-th activation qubits. We depict this quantum circuit in 
\figref{fig:D} (a).
As explained in detail in appendix \ref{sec:details_classical},
the computation of $\mu$ involves the matrix element of $\mK_i$ in the product state $\ket{\psi}$ while
$\sigma^2$ involves that of $\mK_i\mK_{i+1}$.
Due to the structure of $\mD$, these operators act locally on $4$ and $6$ sites respectively as depicted in 
\figref{fig:D} (b)-(c).
\begin{figure}[ht]
\vspace{-0.0cm}
\begin{center}
\mbox{
{\scriptsize
\providecommand{\mynus}[1][.7]{\scalebox{#1}{-}}
\begin{tikzpicture}[scale=.8000000,x=1pt,y=1pt]
\draw[color=black] (-100.000000,-25.000000) node[above] {$\mD=$};
\filldraw[color=white] (7.500000, 0.000000) rectangle (-82.500000, -48.000000);
\draw[color=black] (-75.000000,0.000000) -- (-75.000000,-48.000000);
\draw[color=black] (-75.000000,0.000000) node[above] {$h_0$};
\draw[color=black] (-60.000000,0.000000) -- (-60.000000,-48.000000);
\draw[color=black] (-60.000000,0.000000) node[above] {$W_0$};
\draw[color=black] (-45.000000,0.000000) -- (-45.000000,-48.000000);
\draw[color=black] (-45.000000,0.000000) node[above] {$h_1$};
\draw[color=black] (-30.000000,0.000000) -- (-30.000000,-48.000000);
\draw[color=black] (-30.000000,0.000000) node[above] {$W_1$};
\draw[color=black] (-15.000000,0.000000) -- (-15.000000,-48.000000);
\draw[color=black] (-15.000000,0.000000) node[above] {$h_2$};
\draw[color=black] (-0.000000,0.000000) -- (-0.000000,-48.000000);
\draw[color=black] (-0.000000,0.000000) node[above] {$W_2$};
\draw (-60.000000,-12.000000) -- (-45.000000,-12.000000);
\begin{scope}[color=red]
\begin{scope}
\draw[fill=white] (-52.500000, -12.000000) +(-45.000000:19.091883pt and 8.485281pt) -- +(45.000000:19.091883pt and 8.485281pt) -- +(135.000000:19.091883pt and 8.485281pt) -- +(225.000000:19.091883pt and 8.485281pt) -- cycle;
\clip (-52.500000, -12.000000) +(-45.000000:19.091883pt and 8.485281pt) -- +(45.000000:19.091883pt and 8.485281pt) -- +(135.000000:19.091883pt and 8.485281pt) -- +(225.000000:19.091883pt and 8.485281pt) -- cycle;
\draw (-52.500000, -12.000000) node {$\mP$};
\end{scope}
\end{scope}
\draw (-30.000000,-12.000000) -- (-15.000000,-12.000000);
\begin{scope}[color=red]
\begin{scope}
\draw[fill=white] (-22.500000, -12.000000) +(-45.000000:19.091883pt and 8.485281pt) -- +(45.000000:19.091883pt and 8.485281pt) -- +(135.000000:19.091883pt and 8.485281pt) -- +(225.000000:19.091883pt and 8.485281pt) -- cycle;
\clip (-22.500000, -12.000000) +(-45.000000:19.091883pt and 8.485281pt) -- +(45.000000:19.091883pt and 8.485281pt) -- +(135.000000:19.091883pt and 8.485281pt) -- +(225.000000:19.091883pt and 8.485281pt) -- cycle;
\draw (-22.500000, -12.000000) node {$\mP$};
\end{scope}
\end{scope}
\draw (-75.000000,-36.000000) -- (-60.000000,-36.000000);
\begin{scope}[color=blue]
\begin{scope}
\draw[fill=white] (-67.500000, -36.000000) +(-45.000000:19.091883pt and 8.485281pt) -- +(45.000000:19.091883pt and 8.485281pt) -- +(135.000000:19.091883pt and 8.485281pt) -- +(225.000000:19.091883pt and 8.485281pt) -- cycle;
\clip (-67.500000, -36.000000) +(-45.000000:19.091883pt and 8.485281pt) -- +(45.000000:19.091883pt and 8.485281pt) -- +(135.000000:19.091883pt and 8.485281pt) -- +(225.000000:19.091883pt and 8.485281pt) -- cycle;
\draw (-67.500000, -36.000000) node {$\mQ$};
\end{scope}
\end{scope}
\draw (-45.000000,-36.000000) -- (-30.000000,-36.000000);
\begin{scope}[color=blue]
\begin{scope}
\draw[fill=white] (-37.500000, -36.000000) +(-45.000000:19.091883pt and 8.485281pt) -- +(45.000000:19.091883pt and 8.485281pt) -- +(135.000000:19.091883pt and 8.485281pt) -- +(225.000000:19.091883pt and 8.485281pt) -- cycle;
\clip (-37.500000, -36.000000) +(-45.000000:19.091883pt and 8.485281pt) -- +(45.000000:19.091883pt and 8.485281pt) -- +(135.000000:19.091883pt and 8.485281pt) -- +(225.000000:19.091883pt and 8.485281pt) -- cycle;
\draw (-37.500000, -36.000000) node {$\mQ$};
\end{scope}
\end{scope}
\draw (-15.000000,-36.000000) -- (-0.000000,-36.000000);
\begin{scope}[color=blue]
\begin{scope}
\draw[fill=white] (-7.500000, -36.000000) +(-45.000000:19.091883pt and 8.485281pt) -- +(45.000000:19.091883pt and 8.485281pt) -- +(135.000000:19.091883pt and 8.485281pt) -- +(225.000000:19.091883pt and 8.485281pt) -- cycle;
\clip (-7.500000, -36.000000) +(-45.000000:19.091883pt and 8.485281pt) -- +(45.000000:19.091883pt and 8.485281pt) -- +(135.000000:19.091883pt and 8.485281pt) -- +(225.000000:19.091883pt and 8.485281pt) -- cycle;
\draw (-7.500000, -36.000000) node {$\mQ$};
\end{scope}
\end{scope}
\draw[color=black] (-75.000000,-48.000000) node[below] {$0$};
\draw[color=black] (-60.000000,-48.000000) node[below] {$1$};
\draw[color=black] (-45.000000,-48.000000) node[below] {$2$};
\draw[color=black] (-30.000000,-48.000000) node[below] {$3$};
\draw[color=black] (-15.000000,-48.000000) node[below] {$4$};
\draw[color=black] (-0.000000,-48.000000) node[below] {$5$};
\draw[color=black] (-45.000000,-60.000000) node[below] {(a)};
\end{tikzpicture}
\hspace{0.3cm}
\providecommand{\mynus}[1][.7]{\scalebox{#1}{-}}
\providecommand{\myplus}[1][.7]{\scalebox{#1}{+}}
\begin{tikzpicture}[scale=.9500000,x=1pt,y=1pt]
\draw[color=black] (-60.000000,-70.000000) node[above] {$\mK_i=$};
\filldraw[color=white] (7.500000, 0.000000) rectangle (-52.500000, -120.000000);
\draw[color=black] (-45.000000,0.000000) -- (-45.000000,-120.000000);
\draw[color=black] (-45.000000,0.000000) node[above] {${}$};
\draw[color=black] (-30.000000,0.000000) -- (-30.000000,-120.000000);
\draw[color=black] (-30.000000,0.000000) node[above] {${}$};
\draw[color=black] (-15.000000,0.000000) -- (-15.000000,-120.000000);
\draw[color=black] (-15.000000,0.000000) node[above] {${}$};
\draw[color=black] (-0.000000,0.000000) -- (-0.000000,-120.000000);
\draw[color=black] (-0.000000,0.000000) node[above] {${}$};
\draw (-45.000000,-12.000000) -- (-30.000000,-12.000000);
\begin{scope}[color=red]
\begin{scope}
\draw[fill=white] (-37.500000, -12.000000) +(-45.000000:19.091883pt and 8.485281pt) -- +(45.000000:19.091883pt and 8.485281pt) -- +(135.000000:19.091883pt and 8.485281pt) -- +(225.000000:19.091883pt and 8.485281pt) -- cycle;
\clip (-37.500000, -12.000000) +(-45.000000:19.091883pt and 8.485281pt) -- +(45.000000:19.091883pt and 8.485281pt) -- +(135.000000:19.091883pt and 8.485281pt) -- +(225.000000:19.091883pt and 8.485281pt) -- cycle;
\draw (-37.500000, -12.000000) node {$\mP$};
\end{scope}
\end{scope}
\draw (-15.000000,-12.000000) -- (-0.000000,-12.000000);
\begin{scope}[color=red]
\begin{scope}
\draw[fill=white] (-7.500000, -12.000000) +(-45.000000:19.091883pt and 8.485281pt) -- +(45.000000:19.091883pt and 8.485281pt) -- +(135.000000:19.091883pt and 8.485281pt) -- +(225.000000:19.091883pt and 8.485281pt) -- cycle;
\clip (-7.500000, -12.000000) +(-45.000000:19.091883pt and 8.485281pt) -- +(45.000000:19.091883pt and 8.485281pt) -- +(135.000000:19.091883pt and 8.485281pt) -- +(225.000000:19.091883pt and 8.485281pt) -- cycle;
\draw (-7.500000, -12.000000) node {$\mP$};
\end{scope}
\end{scope}
\draw (-30.000000,-36.000000) -- (-15.000000,-36.000000);
\begin{scope}[color=blue]
\begin{scope}
\draw[fill=white] (-22.500000, -36.000000) +(-45.000000:19.091883pt and 8.485281pt) -- +(45.000000:19.091883pt and 8.485281pt) -- +(135.000000:19.091883pt and 8.485281pt) -- +(225.000000:19.091883pt and 8.485281pt) -- cycle;
\clip (-22.500000, -36.000000) +(-45.000000:19.091883pt and 8.485281pt) -- +(45.000000:19.091883pt and 8.485281pt) -- +(135.000000:19.091883pt and 8.485281pt) -- +(225.000000:19.091883pt and 8.485281pt) -- cycle;
\draw (-22.500000, -36.000000) node {$\mQ$};
\end{scope}
\end{scope}
\draw (-30.000000,-60.000000) -- (-15.000000,-60.000000);
\begin{scope}
\draw[fill=white] (-22.500000, -60.000000) +(-45.000000:19.091883pt and 8.485281pt) -- +(45.000000:19.091883pt and 8.485281pt) -- +(135.000000:19.091883pt and 8.485281pt) -- +(225.000000:19.091883pt and 8.485281pt) -- cycle;
\clip (-22.500000, -60.000000) +(-45.000000:19.091883pt and 8.485281pt) -- +(45.000000:19.091883pt and 8.485281pt) -- +(135.000000:19.091883pt and 8.485281pt) -- +(225.000000:19.091883pt and 8.485281pt) -- cycle;
\draw (-22.500000, -60.000000) node {$\mB\mB$};
\end{scope}
\draw (-30.000000,-84.000000) -- (-15.000000,-84.000000);
\begin{scope}[color=blue]
\begin{scope}
\draw[fill=white] (-22.500000, -84.000000) +(-45.000000:19.091883pt and 8.485281pt) -- +(45.000000:19.091883pt and 8.485281pt) -- +(135.000000:19.091883pt and 8.485281pt) -- +(225.000000:19.091883pt and 8.485281pt) -- cycle;
\clip (-22.500000, -84.000000) +(-45.000000:19.091883pt and 8.485281pt) -- +(45.000000:19.091883pt and 8.485281pt) -- +(135.000000:19.091883pt and 8.485281pt) -- +(225.000000:19.091883pt and 8.485281pt) -- cycle;
\draw (-22.500000, -84.000000) node {$\mQ^{-1}$};
\end{scope}
\end{scope}
\draw (-45.000000,-108.000000) -- (-30.000000,-108.000000);
\begin{scope}[color=red]
\begin{scope}
\draw[fill=white] (-37.500000, -108.000000) +(-45.000000:19.091883pt and 8.485281pt) -- +(45.000000:19.091883pt and 8.485281pt) -- +(135.000000:19.091883pt and 8.485281pt) -- +(225.000000:19.091883pt and 8.485281pt) -- cycle;
\clip (-37.500000, -108.000000) +(-45.000000:19.091883pt and 8.485281pt) -- +(45.000000:19.091883pt and 8.485281pt) -- +(135.000000:19.091883pt and 8.485281pt) -- +(225.000000:19.091883pt and 8.485281pt) -- cycle;
\draw (-37.500000, -108.000000) node {$\mP^{-1}$};
\end{scope}
\end{scope}
\draw (-15.000000,-108.000000) -- (-0.000000,-108.000000);
\begin{scope}[color=red]
\begin{scope}
\draw[fill=white] (-7.500000, -108.000000) +(-45.000000:19.091883pt and 8.485281pt) -- +(45.000000:19.091883pt and 8.485281pt) -- +(135.000000:19.091883pt and 8.485281pt) -- +(225.000000:19.091883pt and 8.485281pt) -- cycle;
\clip (-7.500000, -108.000000) +(-45.000000:19.091883pt and 8.485281pt) -- +(45.000000:19.091883pt and 8.485281pt) -- +(135.000000:19.091883pt and 8.485281pt) -- +(225.000000:19.091883pt and 8.485281pt) -- cycle;
\draw (-7.500000, -108.000000) node {$\mP^{-1}$};
\end{scope}
\end{scope}
\draw[color=black] (-45.000000,-120.000000) node[below] {$2i\mynus1$};
\draw[color=black] (-30.000000,-120.000000) node[below] {$2i$};
\draw[color=black] (-17.000000,-120.000000) node[below] {$2i\myplus1$};
\draw[color=black] (-0.000000,-120.000000) node[below] {$2i\myplus12$};
\draw[color=black] (-25.000000,-130.000000) node[below] {(b)};
\end{tikzpicture}
\hspace{0cm}
\providecommand{\mynus}[1][.7]{\scalebox{#1}{-}}
\providecommand{\myplus}[1][.7]{\scalebox{#1}{+}}
\begin{tikzpicture}[scale=.95000000,x=1pt,y=1pt]
\draw[color=black] (-100.000000,-70.000000) node[above] {$\mK_i\mK_{i+1}=$};
\filldraw[color=white] (7.500000, 0.000000) rectangle (-82.500000, -120.000000);
\draw[color=black] (-75.000000,0.000000) -- (-75.000000,-120.000000);
\draw[color=black] (-75.000000,0.000000) node[above] {${}$};
\draw[color=black] (-60.000000,0.000000) -- (-60.000000,-120.000000);
\draw[color=black] (-60.000000,0.000000) node[above] {${}$};
\draw[color=black] (-45.000000,0.000000) -- (-45.000000,-120.000000);
\draw[color=black] (-45.000000,0.000000) node[above] {${}$};
\draw[color=black] (-30.000000,0.000000) -- (-30.000000,-120.000000);
\draw[color=black] (-30.000000,0.000000) node[above] {${}$};
\draw[color=black] (-15.000000,0.000000) -- (-15.000000,-120.000000);
\draw[color=black] (-15.000000,0.000000) node[above] {${}$};
\draw[color=black] (-0.000000,0.000000) -- (-0.000000,-120.000000);
\draw[color=black] (-0.000000,0.000000) node[above] {${}$};
\draw (-75.000000,-12.000000) -- (-60.000000,-12.000000);
\begin{scope}[color=red]
\begin{scope}
\draw[fill=white] (-67.500000, -12.000000) +(-45.000000:19.091883pt and 8.485281pt) -- +(45.000000:19.091883pt and 8.485281pt) -- +(135.000000:19.091883pt and 8.485281pt) -- +(225.000000:19.091883pt and 8.485281pt) -- cycle;
\clip (-67.500000, -12.000000) +(-45.000000:19.091883pt and 8.485281pt) -- +(45.000000:19.091883pt and 8.485281pt) -- +(135.000000:19.091883pt and 8.485281pt) -- +(225.000000:19.091883pt and 8.485281pt) -- cycle;
\draw (-67.500000, -12.000000) node {$\mP$};
\end{scope}
\end{scope}
\draw (-45.000000,-12.000000) -- (-30.000000,-12.000000);
\begin{scope}[color=red]
\begin{scope}
\draw[fill=white] (-37.500000, -12.000000) +(-45.000000:19.091883pt and 8.485281pt) -- +(45.000000:19.091883pt and 8.485281pt) -- +(135.000000:19.091883pt and 8.485281pt) -- +(225.000000:19.091883pt and 8.485281pt) -- cycle;
\clip (-37.500000, -12.000000) +(-45.000000:19.091883pt and 8.485281pt) -- +(45.000000:19.091883pt and 8.485281pt) -- +(135.000000:19.091883pt and 8.485281pt) -- +(225.000000:19.091883pt and 8.485281pt) -- cycle;
\draw (-37.500000, -12.000000) node {$\mP$};
\end{scope}
\end{scope}
\draw (-15.000000,-12.000000) -- (-0.000000,-12.000000);
\begin{scope}[color=red]
\begin{scope}
\draw[fill=white] (-7.500000, -12.000000) +(-45.000000:19.091883pt and 8.485281pt) -- +(45.000000:19.091883pt and 8.485281pt) -- +(135.000000:19.091883pt and 8.485281pt) -- +(225.000000:19.091883pt and 8.485281pt) -- cycle;
\clip (-7.500000, -12.000000) +(-45.000000:19.091883pt and 8.485281pt) -- +(45.000000:19.091883pt and 8.485281pt) -- +(135.000000:19.091883pt and 8.485281pt) -- +(225.000000:19.091883pt and 8.485281pt) -- cycle;
\draw (-7.500000, -12.000000) node {$\mP$};
\end{scope}
\end{scope}
\draw (-60.000000,-36.000000) -- (-45.000000,-36.000000);
\begin{scope}[color=blue]
\begin{scope}
\draw[fill=white] (-52.500000, -36.000000) +(-45.000000:19.091883pt and 8.485281pt) -- +(45.000000:19.091883pt and 8.485281pt) -- +(135.000000:19.091883pt and 8.485281pt) -- +(225.000000:19.091883pt and 8.485281pt) -- cycle;
\clip (-52.500000, -36.000000) +(-45.000000:19.091883pt and 8.485281pt) -- +(45.000000:19.091883pt and 8.485281pt) -- +(135.000000:19.091883pt and 8.485281pt) -- +(225.000000:19.091883pt and 8.485281pt) -- cycle;
\draw (-52.500000, -36.000000) node {$\mQ$};
\end{scope}
\end{scope}
\draw (-30.000000,-36.000000) -- (-15.000000,-36.000000);
\begin{scope}[color=blue]
\begin{scope}
\draw[fill=white] (-22.500000, -36.000000) +(-45.000000:19.091883pt and 8.485281pt) -- +(45.000000:19.091883pt and 8.485281pt) -- +(135.000000:19.091883pt and 8.485281pt) -- +(225.000000:19.091883pt and 8.485281pt) -- cycle;
\clip (-22.500000, -36.000000) +(-45.000000:19.091883pt and 8.485281pt) -- +(45.000000:19.091883pt and 8.485281pt) -- +(135.000000:19.091883pt and 8.485281pt) -- +(225.000000:19.091883pt and 8.485281pt) -- cycle;
\draw (-22.500000, -36.000000) node {$\mQ$};
\end{scope}
\end{scope}
\draw (-60.000000,-60.000000) -- (-45.000000,-60.000000);
\begin{scope}
\draw[fill=white] (-52.500000, -60.000000) +(-45.000000:19.091883pt and 8.485281pt) -- +(45.000000:19.091883pt and 8.485281pt) -- +(135.000000:19.091883pt and 8.485281pt) -- +(225.000000:19.091883pt and 8.485281pt) -- cycle;
\clip (-52.500000, -60.000000) +(-45.000000:19.091883pt and 8.485281pt) -- +(45.000000:19.091883pt and 8.485281pt) -- +(135.000000:19.091883pt and 8.485281pt) -- +(225.000000:19.091883pt and 8.485281pt) -- cycle;
\draw (-52.500000, -60.000000) node {$\mB\mB$};
\end{scope}
\draw (-30.000000,-60.000000) -- (-15.000000,-60.000000);
\begin{scope}
\draw[fill=white] (-22.500000, -60.000000) +(-45.000000:19.091883pt and 8.485281pt) -- +(45.000000:19.091883pt and 8.485281pt) -- +(135.000000:19.091883pt and 8.485281pt) -- +(225.000000:19.091883pt and 8.485281pt) -- cycle;
\clip (-22.500000, -60.000000) +(-45.000000:19.091883pt and 8.485281pt) -- +(45.000000:19.091883pt and 8.485281pt) -- +(135.000000:19.091883pt and 8.485281pt) -- +(225.000000:19.091883pt and 8.485281pt) -- cycle;
\draw (-22.500000, -60.000000) node {$\mB\mB$};
\end{scope}
\draw (-60.000000,-84.000000) -- (-45.000000,-84.000000);
\begin{scope}[color=blue]
\begin{scope}
\draw[fill=white] (-52.500000, -84.000000) +(-45.000000:19.091883pt and 8.485281pt) -- +(45.000000:19.091883pt and 8.485281pt) -- +(135.000000:19.091883pt and 8.485281pt) -- +(225.000000:19.091883pt and 8.485281pt) -- cycle;
\clip (-52.500000, -84.000000) +(-45.000000:19.091883pt and 8.485281pt) -- +(45.000000:19.091883pt and 8.485281pt) -- +(135.000000:19.091883pt and 8.485281pt) -- +(225.000000:19.091883pt and 8.485281pt) -- cycle;
\draw (-52.500000, -84.000000) node {$\mQ^{-1}$};
\end{scope}
\end{scope}
\draw (-30.000000,-84.000000) -- (-15.000000,-84.000000);
\begin{scope}[color=blue]
\begin{scope}
\draw[fill=white] (-22.500000, -84.000000) +(-45.000000:19.091883pt and 8.485281pt) -- +(45.000000:19.091883pt and 8.485281pt) -- +(135.000000:19.091883pt and 8.485281pt) -- +(225.000000:19.091883pt and 8.485281pt) -- cycle;
\clip (-22.500000, -84.000000) +(-45.000000:19.091883pt and 8.485281pt) -- +(45.000000:19.091883pt and 8.485281pt) -- +(135.000000:19.091883pt and 8.485281pt) -- +(225.000000:19.091883pt and 8.485281pt) -- cycle;
\draw (-22.500000, -84.000000) node {$\mQ^{-1}$};
\end{scope}
\end{scope}
\draw (-75.000000,-108.000000) -- (-60.000000,-108.000000);
\begin{scope}[color=red]
\begin{scope}
\draw[fill=white] (-67.500000, -108.000000) +(-45.000000:19.091883pt and 8.485281pt) -- +(45.000000:19.091883pt and 8.485281pt) -- +(135.000000:19.091883pt and 8.485281pt) -- +(225.000000:19.091883pt and 8.485281pt) -- cycle;
\clip (-67.500000, -108.000000) +(-45.000000:19.091883pt and 8.485281pt) -- +(45.000000:19.091883pt and 8.485281pt) -- +(135.000000:19.091883pt and 8.485281pt) -- +(225.000000:19.091883pt and 8.485281pt) -- cycle;
\draw (-67.500000, -108.000000) node {$\mP^{-1}$};
\end{scope}
\end{scope}
\draw (-45.000000,-108.000000) -- (-30.000000,-108.000000);
\begin{scope}[color=red]
\begin{scope}
\draw[fill=white] (-37.500000, -108.000000) +(-45.000000:19.091883pt and 8.485281pt) -- +(45.000000:19.091883pt and 8.485281pt) -- +(135.000000:19.091883pt and 8.485281pt) -- +(225.000000:19.091883pt and 8.485281pt) -- cycle;
\clip (-37.500000, -108.000000) +(-45.000000:19.091883pt and 8.485281pt) -- +(45.000000:19.091883pt and 8.485281pt) -- +(135.000000:19.091883pt and 8.485281pt) -- +(225.000000:19.091883pt and 8.485281pt) -- cycle;
\draw (-37.500000, -108.000000) node {$\mP^{-1}$};
\end{scope}
\end{scope}
\draw (-15.000000,-108.000000) -- (-0.000000,-108.000000);
\begin{scope}[color=red]
\begin{scope}
\draw[fill=white] (-7.500000, -108.000000) +(-45.000000:19.091883pt and 8.485281pt) -- +(45.000000:19.091883pt and 8.485281pt) -- +(135.000000:19.091883pt and 8.485281pt) -- +(225.000000:19.091883pt and 8.485281pt) -- cycle;
\clip (-7.500000, -108.000000) +(-45.000000:19.091883pt and 8.485281pt) -- +(45.000000:19.091883pt and 8.485281pt) -- +(135.000000:19.091883pt and 8.485281pt) -- +(225.000000:19.091883pt and 8.485281pt) -- cycle;
\draw (-7.500000, -108.000000) node {$\mP^{-1}$};
\end{scope}
\end{scope}
\draw[color=black] (-75.000000,-120.000000) node[below] {$2i\mynus1$};
\draw[color=black] (-60.000000,-120.000000) node[below] {$2i$};
\draw[color=black] (-45.000000,-120.000000) node[below] {$2i\myplus1$};
\draw[color=black] (-30.000000,-120.000000) node[below] {$2i\myplus2$};
\draw[color=black] (-15.000000,-120.000000) node[below] {$2i\myplus3$};
\draw[color=black] (-0.000000,-120.000000) node[below] {$2i\myplus4$};
\draw[color=black] (-37.000000,-130.000000) node[below] {(c)};
\end{tikzpicture}
}}
\end{center}
\vspace{-0.5cm}
\caption{(a) The entangling circuit $\mD$ for $N=3$. (b) $\mK_i$ entering the computation of $\mu$.
(c) $\mK_i\mK_{i+1}$ entering the computation of $\sigma^2$. Indices on $\mP,\mQ,\mB$ are omitted for clarity. Time flows downwards.}
\label{fig:D}
\end{figure}
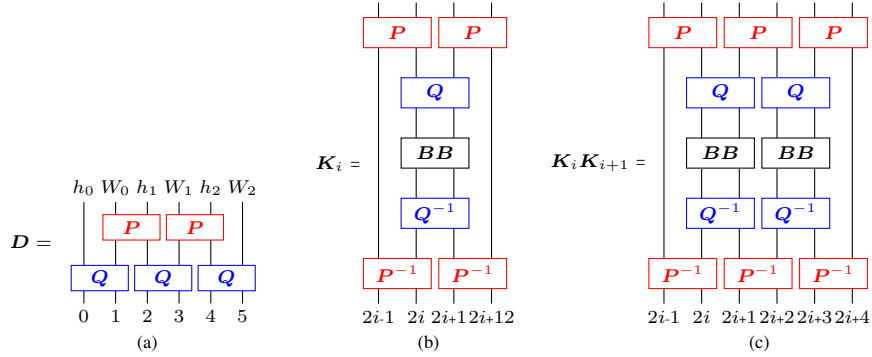
This implies that the computation of \eqref{eq:mu_sigma}, and so of the full layer, 
can be done in $O(N)$ and easily parallelized.
Appendix \ref{sec:details_classical} contains more details on the complexity, while Procedure \ref{algc} describes the algorithm for the classical simulation discussed here.


\begin{algorithm} 
\caption{Quantum deformed layer. QPE$_j$($\mU, {\cal I}$) is quantum phase estimation for a unitary $\mU$ acting on the set ${\cal I}$ of activation qubits and the $j$-th weights/ancilla qubits. $\mH_j$ is in \eqref{eq:H}.} 
\label{algq}
\begin{algorithmic}
    \REQUIRE 
    $\{ q_{ij} \}_{i=0,\dots,N-1}^{j=0,\dots,M-1}$,  $\ket{\psi}$, ${\cal I}$, $\mD$, $t$
    \ENSURE $\ket{\psi}$
    \FOR{$j=0$ \TO $M-1$ } 
        \STATE $\ket{\psi}_{\mW_{j,:}}
            \leftarrow
            \bigotimes_{i=1}^N
            \left[\sqrt{q_{ji}} \ket{0} +     \sqrt{1-q_{ji}} \ket{1}\right]
            $
        \STATE $\ket{\psi}\leftarrow \ket{0}^{\otimes t}\otimes \ket{\psi}\otimes \ket{\psi}_{\mW_{j,:}}$
        \STATE $\mU \leftarrow \mD {\rm e}^{\frac{2\pi i}{N+1} \mH_j} \mD^{-1}$
        \COMMENT{This requires to approximate the unitary with quantum gates}
        \STATE $\ket{\psi}\leftarrow \text{QPE}_j(\mU, {\cal I})\ket{\psi}$
    \ENDFOR
\end{algorithmic}
\end{algorithm}

\begin{algorithm} 
\caption{Classical simulation of a quantum deformed layer with $N$ ($M$) inputs (outputs).}
\label{algc}
\begin{algorithmic} 
    \REQUIRE $\{ q_{ij} \}_{i=0,\dots,N-1}^{j=0,\dots,M-1}, 
    \{ p_i \}_{i=0,\dots,N-1}, 
    \mP = \{ \mP_{2i-1,2i}^j \}_{i=1,\dots,N-1}^{j=0,\dots,M-1}, 
    \mQ = \{ \mQ_{2i,2i+1}^j \}_{i=0,\dots,N-1}^{j=0,\dots,M-1}$
    \ENSURE $\{ p'_i \}_{i=1,\dots,M}$
    \FOR{$j=0$ \TO $M-1$ } 
        \FOR{$i=0$ \TO $N-1$ } 
        \STATE $\psi_{2i}
        \leftarrow
        \left[\sqrt{p_i} , \sqrt{1-p_i}\right]
        $
        \STATE $\psi_{2i+1}
        \leftarrow
        \left[\sqrt{q_{ij}} , \sqrt{1-q_{ij}}\right]$
        \ENDFOR
        \FOR{$i=0$ \TO $N-1$ } 
        \STATE $\mu_i
        \leftarrow$ computeMu($i, \psi, \mP, \mQ$)
        \COMMENT{This implements \eqref{eq:mu} of appendix \ref{sec:details_classical}}
        \STATE $\gamma_{i,i+1}
        \leftarrow$ computeGamma($i, \psi, \mP, \mQ$)
        \COMMENT{This implements \eqref{eq:gamma} of appendix \ref{sec:details_classical}}
        \ENDFOR
        \STATE $\mu\leftarrow \sum_{i=0}^{N-1} \mu_i$
        \STATE $\sigma^2\leftarrow 2 \sum_{i=0}^{N-2}
        (
        \gamma_{i,i+1}
        -
        \mu_i \mu_{i+1}
        )
        +
        \sum_{i=0}^{N-1}(\mu_i - \mu_i^2)$
        \STATE
        $p'_j
        \leftarrow
        \Phi\left(- \tfrac{2\mu-N}{2\sqrt{\sigma^2}} \right)$
    \ENDFOR
\end{algorithmic}
\end{algorithm}

\newpage

\subsection{Experiments}
\label{sec:exp}

\begin{wraptable}{r}{7cm}
\caption{Test accuracies for MNIST and Fashion MNIST.
With the notation $cKsS-C$ to indicate a conv2d layer with $C$ filters of size $[K,K]$ and stride $S$, and $dN$ for a dense layer with $N$ output neurons, the architectures (Arch.) are A: d10; B: c3s2-8, c3s2-16, d10; C: c3s2-32, c3s2-64, d10. The deformations are:
[/]: $\mP^j_{i,i+1}\!=\!\mQ^j_{i,i+1}\!=\!{\bf 1}$ (baseline \citep{pbnn});
[PQ]: $\mP^j_{i,i+1},\mQ^j_{i,i+1}$ generic;
[Q]: $\mP^j_{i,i+1}\!=\!{\bf 1},\mQ^j_{i,i+1}$ generic.
}
\begin{center}
\tabcolsep=0.10cm
\begin{tabular}{cccc}
\hline
Arch. & Deformation & MNIST & Fashion MNIST\\
\hline
A
& [/] & 91.1 & 84.2\\
& [PQ] & {\bf 94.3} & {\bf 86.8}\\
& [Q] & 91.6 & 85.1\\
\hline
B
& [/, /, /]  & 96.6 & 87.5\\
& [PQ, /, /] & {\bf 97.6} & {\bf 88.1}\\
& [Q, /, /] & 96.8 & 87.8\\
\hline
C
& [/, /, /] &  98.1 & 89.3\\
& [PQ, /, /] & {\bf 98.3} & {\bf 89.6}\\
& [Q, /, /] & 98.3 & 89.5\\
\hline
\end{tabular}
\end{center}
\label{tab:results}
\end{wraptable}

We present experiments for the model of the previous section. At each layer, $q_{ij}$ and $\mD_j$ are learnable. They are optimized to minimize the loss of \eqref{eq:objective} where following \citep{pbnn, shayer} we take ${\cal R} = \beta \sum_{\ell,i,j} q_{ij}^{(\ell)}(1-q_{ij}^{(\ell)})$, and ${\cal R}'$ is the $L_2$ regularization loss of the parameters of $\mD_j$. ${\cal L}$ coincides with \eqref{eq:obj_vo}.
We implemented and trained several architectures with different deformations. 
Table \ref{tab:results} contains results for two standard image datasets, MNIST and Fashion MNIST. Details of the experiments are in appendix \ref{sec:details_experiments}.
The classical baseline is based on \citep{pbnn}, but we use fewer layers to make the simulation of the deformation cheaper and use no batch norm, and no max pooling.

The general deformation ([PQ]) performs best in all cases. In the simplest case of a single dense layer (A), the gain is $+3.2\%$ for MNIST and $+2.6\%$ for Fashion MNIST on test accuracy. For convnets, we could only simulate a single deformed layer due to computational issues and the gain is around or less than $1\%$. We expect that deforming all layers will give a greater boost as the improvements diminish with decreasing the ratio of deformation parameters over classical parameters ($q_{ij}$).
The increase in accuracy comes at the expense of more parameters. In appendix \ref{sec:details_experiments} we present additional results showing that quantum models can still deliver modest accuracy improvement w.r.t.~convolutional networks with the same number of parameters.

\section{Conclusions}

In this work we made the following main contributions:
1) we introduced quantum deformed neural networks and identified potential speedups by running these models on a quantum computer; 2) we devised classically efficient algorithms to train the networks for low entanglement designs of the quantum circuits; 3) for the first time in the literature, we simulated the quantum neural networks on real world data sizes obtaining good accuracy, and showed modest gains due to the quantum deformations.
Running these models on a quantum computer will allow one to explore efficiently more general deformations, in particular those that cannot be approximated by the central limit theorem when the Hamiltonians will be sums of non-commuting operators. 
Another interesting future direction is to incorporate batch normalization and pooling layers in quantum neural networks.

An outstanding question in quantum machine learning is to find quantum advantages for classical machine learning tasks. 
The class of known problems for which a quantum learner can have a provably exponential advantage over a classical learner is small at the moment \cite{2020arXiv201002174L}, and some problems that are classically hard to compute can be predicted easily with classical machine learning \cite{huang2020power}.
The approach presented here is the next step in a series of papers that tries to benchmark quantum neural networks empirically, e.g.~\cite{farhi_neven, huggins, Grant_2019, grant2018hierarchical, bausch2020recurrent}. We are the first to show that towards the limit of entangling circuits the quantum inspired architecture does improve relative to the classical one for real world data sizes.

\bibliography{iclr2021_conference}

\begin{thebibliography}{37}
\providecommand{\natexlab}[1]{#1}
\providecommand{\url}[1]{\texttt{#1}}
\expandafter\ifx\csname urlstyle\endcsname\relax
  \providecommand{\doi}[1]{doi: #1}\else
  \providecommand{\doi}{doi: \begingroup \urlstyle{rm}\Url}\fi

\bibitem[Aaronson \& Chen(2016)Aaronson and
  Chen]{aaronson2016complexitytheoretic}
Scott Aaronson and Lijie Chen.
\newblock Complexity-theoretic foundations of quantum supremacy experiments,
  2016.

\bibitem[{Allcock} et~al.(2018){Allcock}, {Hsieh}, {Kerenidis}, and
  {Zhang}]{allcock}
Jonathan {Allcock}, Chang-Yu {Hsieh}, Iordanis {Kerenidis}, and Shengyu
  {Zhang}.
\newblock {Quantum algorithms for feedforward neural networks}.
\newblock \emph{arXiv e-prints}, art. arXiv:1812.03089, December 2018.

\bibitem[Arute et~al.(2019)Arute, Arya, Babbush, Bacon, Bardin, Barends,
  Biswas, Boixo, Brandao, Buell, et~al.]{arute2019quantum}
Frank Arute, Kunal Arya, Ryan Babbush, Dave Bacon, Joseph~C Bardin, Rami
  Barends, Rupak Biswas, Sergio Boixo, Fernando~GSL Brandao, David~A Buell,
  et~al.
\newblock Quantum supremacy using a programmable superconducting processor.
\newblock \emph{Nature}, 574\penalty0 (7779):\penalty0 505--510, 2019.

\bibitem[Bausch(2020)]{bausch2020recurrent}
Johannes Bausch.
\newblock Recurrent quantum neural networks.
\newblock \emph{Advances in Neural Information Processing Systems}, 33, 2020.

\bibitem[{Beer} et~al.(2019){Beer}, {Bondarenko}, {Farrelly}, {Osborne},
  {Salzmann}, and {Wolf}]{beer}
Kerstin {Beer}, Dmytro {Bondarenko}, Terry {Farrelly}, Tobias~J. {Osborne},
  Robert {Salzmann}, and Ramona {Wolf}.
\newblock {Efficient Learning for Deep Quantum Neural Networks}.
\newblock \emph{arXiv e-prints}, art. arXiv:1902.10445, February 2019.

\bibitem[Biamonte et~al.(2017)Biamonte, Wittek, Pancotti, Rebentrost, Wiebe,
  and Lloyd]{Biamonte}
Jacob Biamonte, Peter Wittek, Nicola Pancotti, Patrick Rebentrost, Nathan
  Wiebe, and Seth Lloyd.
\newblock Quantum machine learning.
\newblock \emph{Nature}, 549\penalty0 (7671):\penalty0 195--202, 2017.

\bibitem[{Cao} et~al.(2017){Cao}, {Giacomo Guerreschi}, and
  {Aspuru-Guzik}]{quantum_neuron}
Yudong {Cao}, Gian {Giacomo Guerreschi}, and Al{\'a}n {Aspuru-Guzik}.
\newblock {Quantum Neuron: an elementary building block for machine learning on
  quantum computers}.
\newblock \emph{arXiv e-prints}, art. arXiv:1711.11240, November 2017.

\bibitem[{Chen} et~al.(2018){Chen}, {Rubanova}, {Bettencourt}, and
  {Duvenaud}]{Duvenaud}
Ricky T.~Q. {Chen}, Yulia {Rubanova}, Jesse {Bettencourt}, and David
  {Duvenaud}.
\newblock {Neural Ordinary Differential Equations}.
\newblock \emph{arXiv e-prints}, art. arXiv:1806.07366, Jun 2018.

\bibitem[Cheng et~al.(2020)Cheng, Wang, and Zhang]{cheng2020supervised}
Song Cheng, Lei Wang, and Pan Zhang.
\newblock Supervised learning with projected entangled pair states, 2020.

\bibitem[{Ciliberto} et~al.(2018){Ciliberto}, {Herbster}, {Ialongo}, {Pontil},
  {Rocchetto}, {Severini}, and {Wossnig}]{ciliberto}
Carlo {Ciliberto}, Mark {Herbster}, Alessand ro~Davide {Ialongo}, Massimiliano
  {Pontil}, Andrea {Rocchetto}, Simone {Severini}, and Leonard {Wossnig}.
\newblock {Quantum machine learning: a classical perspective}.
\newblock \emph{Proceedings of the Royal Society of London Series A},
  474\penalty0 (2209):\penalty0 20170551, January 2018.
\newblock \doi{10.1098/rspa.2017.0551}.

\bibitem[{Cong} et~al.(2019){Cong}, {Choi}, and {Lukin}]{qcnn}
Iris {Cong}, Soonwon {Choi}, and Mikhail~D. {Lukin}.
\newblock {Quantum convolutional neural networks}.
\newblock \emph{Nature Physics}, 15\penalty0 (12):\penalty0 1273--1278, August
  2019.
\newblock \doi{10.1038/s41567-019-0648-8}.

\bibitem[{Farhi} \& {Neven}(2018){Farhi} and {Neven}]{farhi_neven}
Edward {Farhi} and Hartmut {Neven}.
\newblock {Classification with Quantum Neural Networks on Near Term
  Processors}.
\newblock \emph{arXiv e-prints}, art. arXiv:1802.06002, February 2018.

\bibitem[{Fuchs} \& {Schack}(2013){Fuchs} and {Schack}]{Fuchsetal}
Christopher~A. {Fuchs} and R{\"u}diger {Schack}.
\newblock {Quantum-Bayesian coherence}.
\newblock \emph{Reviews of Modern Physics}, 85\penalty0 (4):\penalty0
  1693--1715, Oct 2013.
\newblock \doi{10.1103/RevModPhys.85.1693}.

\bibitem[Grant et~al.(2018)Grant, Benedetti, Cao, Hallam, Lockhart, Stojevic,
  Green, and Severini]{grant2018hierarchical}
Edward Grant, Marcello Benedetti, Shuxiang Cao, Andrew Hallam, Joshua Lockhart,
  Vid Stojevic, Andrew~G Green, and Simone Severini.
\newblock Hierarchical quantum classifiers.
\newblock \emph{npj Quantum Information}, 4\penalty0 (1):\penalty0 1--8, 2018.

\bibitem[Grant et~al.(2019)Grant, Wossnig, Ostaszewski, and
  Benedetti]{Grant_2019}
Edward Grant, Leonard Wossnig, Mateusz Ostaszewski, and Marcello Benedetti.
\newblock An initialization strategy for addressing barren plateaus in
  parametrized quantum circuits.
\newblock \emph{Quantum}, 3:\penalty0 214, Dec 2019.
\newblock ISSN 2521-327X.
\newblock \doi{10.22331/q-2019-12-09-214}.
\newblock URL \url{http://dx.doi.org/10.22331/q-2019-12-09-214}.

\bibitem[Huang et~al.(2020)Huang, Broughton, Mohseni, Babbush, Boixo, Neven,
  and McClean]{huang2020power}
Hsin-Yuan Huang, Michael Broughton, Masoud Mohseni, Ryan Babbush, Sergio Boixo,
  Hartmut Neven, and Jarrod~R. McClean.
\newblock Power of data in quantum machine learning, 2020.

\bibitem[{Huggins} et~al.(2019){Huggins}, {Patil}, {Mitchell}, {Whaley}, and
  {Miles Stoudenmire}]{huggins}
William {Huggins}, Piyush {Patil}, Bradley {Mitchell}, K.~Birgitta {Whaley},
  and E.~{Miles Stoudenmire}.
\newblock {Towards quantum machine learning with tensor networks}.
\newblock \emph{Quantum Science and Technology}, 4\penalty0 (2):\penalty0
  024001, Apr 2019.
\newblock \doi{10.1088/2058-9565/aaea94}.

\bibitem[Kerenidis et~al.(2019)Kerenidis, Landman, and
  Prakash]{kerenidis2019quantum}
Iordanis Kerenidis, Jonas Landman, and Anupam Prakash.
\newblock Quantum algorithms for deep convolutional neural networks, 2019.

\bibitem[Kingma et~al.(2015)Kingma, Salimans, and
  Welling]{kingma2015variational}
Diederik~P. Kingma, Tim Salimans, and Max Welling.
\newblock Variational dropout and the local reparameterization trick, 2015.

\bibitem[Levin \& Peres(2017)Levin and Peres]{levin2017markov}
David~A Levin and Yuval Peres.
\newblock \emph{Markov chains and mixing times}, volume 107.
\newblock American Mathematical Soc., 2017.

\bibitem[{Levine} et~al.(2017){Levine}, {Yakira}, {Cohen}, and
  {Shashua}]{convac}
Yoav {Levine}, David {Yakira}, Nadav {Cohen}, and Amnon {Shashua}.
\newblock {Deep Learning and Quantum Entanglement: Fundamental Connections with
  Implications to Network Design}.
\newblock \emph{arXiv e-prints}, art. arXiv:1704.01552, April 2017.

\bibitem[Levine et~al.(2019)Levine, Sharir, Cohen, and Shashua]{convac2}
Yoav Levine, Or~Sharir, Nadav Cohen, and Amnon Shashua.
\newblock Quantum entanglement in deep learning architectures.
\newblock \emph{Physical review letters}, 122\penalty0 (6):\penalty0 065301,
  2019.

\bibitem[{Liu} et~al.(2017){Liu}, {Ran}, {Wittek}, {Peng}, {Bl{\'a}zquez
  Garc{\'\i}a}, {Su}, and {Lewenstein}]{Liu}
Ding {Liu}, Shi-Ju {Ran}, Peter {Wittek}, Cheng {Peng}, Raul {Bl{\'a}zquez
  Garc{\'\i}a}, Gang {Su}, and Maciej {Lewenstein}.
\newblock {Machine Learning by Unitary Tensor Network of Hierarchical Tree
  Structure}.
\newblock \emph{arXiv e-prints}, art. arXiv:1710.04833, October 2017.

\bibitem[{Liu} et~al.(2020){Liu}, {Arunachalam}, and
  {Temme}]{2020arXiv201002174L}
Yunchao {Liu}, Srinivasan {Arunachalam}, and Kristan {Temme}.
\newblock {A rigorous and robust quantum speed-up in supervised machine
  learning}.
\newblock \emph{arXiv e-prints}, art. arXiv:2010.02174, October 2020.

\bibitem[{Miles Stoudenmire} \& {Schwab}(2016){Miles Stoudenmire} and
  {Schwab}]{stoudenmire}
E.~{Miles Stoudenmire} and David~J. {Schwab}.
\newblock {Supervised Learning with Quantum-Inspired Tensor Networks}.
\newblock \emph{arXiv e-prints}, art. arXiv:1605.05775, May 2016.

\bibitem[Nielsen \& Chuang(2000)Nielsen and Chuang]{NielsenChuang}
M.E. Nielsen and I.L. Chuang.
\newblock \emph{Quantum Computation and Quantum Information}.
\newblock Cambridge Series on Information and the Natural Sciences. Cambridge
  University Press, 2000.
\newblock ISBN 9780521635035.
\newblock URL \url{https://books.google.co.uk/books?id=aai-P4V9GJ8C}.

\bibitem[Peters \& Welling(2018)Peters and Welling]{pbnn}
Jorn W.~T. Peters and Max Welling.
\newblock Probabilistic binary neural networks, 2018.

\bibitem[Schuld et~al.(2015)Schuld, Sinayskiy, and Petruccione]{Schuld_2015}
Maria Schuld, Ilya Sinayskiy, and Francesco Petruccione.
\newblock Simulating a perceptron on a quantum computer.
\newblock \emph{Physics Letters A}, 379\penalty0 (7):\penalty0 660–663, Mar
  2015.
\newblock ISSN 0375-9601.
\newblock \doi{10.1016/j.physleta.2014.11.061}.
\newblock URL \url{http://dx.doi.org/10.1016/j.physleta.2014.11.061}.

\bibitem[Shayer et~al.(2017)Shayer, Levi, and Fetaya]{shayer}
Oran Shayer, Dan Levi, and Ethan Fetaya.
\newblock Learning discrete weights using the local reparameterization trick.
\newblock \emph{arXiv preprint arXiv:1710.07739}, 2017.

\bibitem[Staines \& Barber(2012)Staines and Barber]{staines2012variational}
Joe Staines and David Barber.
\newblock Variational optimization, 2012.

\bibitem[;t~Hooft(2016)]{hooft2016cellular}
G.~;t~Hooft.
\newblock \emph{The Cellular Automaton Interpretation of Quantum Mechanics}.
\newblock Fundamental Theories of Physics. Springer International Publishing,
  2016.
\newblock ISBN 9783319412856.
\newblock URL \url{https://books.google.co.uk/books?id=ctlCDwAAQBAJ}.

\bibitem[Tang(2019)]{tang}
Ewin Tang.
\newblock A quantum-inspired classical algorithm for recommendation systems.
\newblock In \emph{Proceedings of the 51st Annual ACM SIGACT Symposium on
  Theory of Computing}, STOC 2019, pp.\  217–228, New York, NY, USA, 2019.
  Association for Computing Machinery.
\newblock ISBN 9781450367059.
\newblock \doi{10.1145/3313276.3316310}.
\newblock URL \url{https://doi.org/10.1145/3313276.3316310}.

\bibitem[{Verdon} et~al.(2018){Verdon}, {Pye}, and {Broughton}]{baqprop}
Guillaume {Verdon}, Jason {Pye}, and Michael {Broughton}.
\newblock {A Universal Training Algorithm for Quantum Deep Learning}.
\newblock \emph{arXiv e-prints}, art. arXiv:1806.09729, June 2018.

\bibitem[Vidal(2003)]{vidal2003efficient}
Guifr{\'e} Vidal.
\newblock Efficient classical simulation of slightly entangled quantum
  computations.
\newblock \emph{Physical review letters}, 91\penalty0 (14):\penalty0 147902,
  2003.

\bibitem[{Wiebe} et~al.(2014){Wiebe}, {Kapoor}, and {Svore}]{quantumDL}
Nathan {Wiebe}, Ashish {Kapoor}, and Krysta~M. {Svore}.
\newblock {Quantum Deep Learning}.
\newblock \emph{arXiv e-prints}, art. arXiv:1412.3489, December 2014.

\bibitem[Zhao et~al.(2019{\natexlab{a}})Zhao, Fitzsimons, and
  Fitzsimons]{quantumGP2}
Zhikuan Zhao, Jack~K. Fitzsimons, and Joseph~F. Fitzsimons.
\newblock Quantum-assisted gaussian process regression.
\newblock \emph{Phys. Rev. A}, 99:\penalty0 052331, May 2019{\natexlab{a}}.
\newblock \doi{10.1103/PhysRevA.99.052331}.
\newblock URL \url{https://link.aps.org/doi/10.1103/PhysRevA.99.052331}.

\bibitem[Zhao et~al.(2019{\natexlab{b}})Zhao, Fitzsimons, Osborne, Roberts, and
  Fitzsimons]{quantumGP1}
Zhikuan Zhao, Jack~K. Fitzsimons, Michael~A. Osborne, Stephen~J. Roberts, and
  Joseph~F. Fitzsimons.
\newblock Quantum algorithms for training gaussian processes.
\newblock \emph{Phys. Rev. A}, 100:\penalty0 012304, Jul 2019{\natexlab{b}}.
\newblock \doi{10.1103/PhysRevA.100.012304}.
\newblock URL \url{https://link.aps.org/doi/10.1103/PhysRevA.100.012304}.

\end{thebibliography}
\bibliographystyle{iclr2021_conference}


\newpage
\appendix

\section{Variational Bayes}
\label{sec:elbo_vo}

In this section we review variational Bayes.

Given $M$ input/output pairs $\mX = (\bm{x}^1,\dots, \bm{x}^M), \mY = (\bm{y}^1,\dots,\bm{y}^M)$, and a prior over weights $p(\mW)$,
in a Bayesian approach we are interested in computing the posterior $p(\mW | \mX, \mY) = p(\mY |\mX, \mW ) p(\mW) / p(\mY|\mX)$, which is used to make a prediction on a test point $\bm{x}_*$: $p(\bm{y}|\bm{x}_*) = \mathbb{E}_{p(\mW | \mX, \mY)}[p(\bm{y}|\bm{x}_*,\mW)]$.
However the posterior computation involves an intractable denominator. We proceed using variational Bayes and we introduce an approximate posterior $q_\theta(\mW)$ which depends on hyperparameters $\theta$, chosen to maximize the evidence $p(\mY|\mX)$ lower bound (ELBO) objective:
\begin{align}
    \label{eq:ELBO}
    \text{ELBO}
    =
    \mathbb{E}_{q_\theta(\mW)}[\log p(\mY |\mX, \mW )]
    -
    \text{KL}(q_\theta(\mW) || p(\mW))
    \le 
    p(\mY|\mX)
    \,.
\end{align}
The resulting approximate posterior is then used for a prediction:
$q(\bm{y}|\bm{x}_*) := \mathbb{E}_{q_\theta(\mW)}[p(\bm{y}|\bm{x}_*,\mW)]$.
Maximizing the gap in variational Bayes is equivalent to minimizing the KL between the approximate and the true posterior \cite{kingma2015variational}: 
\begin{align}
    \text{ELBO} - \log p(\mY|\mX)
    =
    -\text{KL}(q_\theta(\mW) || p(\mW|\mX, \mY))
    \,.
\end{align}


\section{Review of quantum mechanics}
\label{sec:qm}

States in quantum mechanics (QM) are represented as abstract vectors in a Hilbert space ${\cal H}$, denoted as $\ket{\psi}$. Here, we will only be concerned with qubits which can have two possible states $\ket{0}$ and $\ket{1}$. These states are defined relative to a particular frame of reference, e.g. spin values measured along the z-axis. More generally, a qubit is described as a complex linear combination (or superposition) of up and down z-spins: $\ket{\psi}=\alpha\ket{0} + \beta\ket{1}$. 

$N$ \emph{disentangled} qubits are described by a product state $\ket{\psi}=\prod_{i=1}^{N}\ket{\psi_i}$. Under interactions qubits entangle with each other. This means that the wave function is now a complex linear combination of an exponential number of $2^N$ terms. We can write this as $\ket{\psi'}=U\ket{\psi}$ where $U$ is a unitary matrix in a $2^N\times 2^N$ dimensional space. This entangled state is still 'pure' in the sense that there is nothing more to learn about it, i.e. its quantum entropy is zero and it represents maximal information about the system. However, in QM that does not mean our knowledge of the system is complete. 

Time evolution in QM is described by a unitary transformation, $\ket{\psi(t)}=\mU(t,0)\ket{\psi(0)}$ with $\mU(t,0)=e^{i\mH t}$ where H is the Hamiltonian, a Hermitian operator. Note that time evolution entangles qubits. We will use time evolution to map an input state to an output state as a layer in a NN, not unlike a neural ODE \cite{Duvenaud}. 

Measurements in QM are nothing else than projecting the state onto the eigenbasis of a symmetric positive definite operator $\mA$. The quantum system collapses into a particular state with a probability given by Born's rule: $p_i=|\bra{\phi_i}\ket{\psi}|^2$ where $\{\ket{\phi_i}\}$ are the orthonormal eigenvectors of $\mA$. 

We will also need to describe ``mixed states''. A mixed state is a classical mixture of a number of pure quantum states. Probabilities in this mixture encode our uncertainty about what quantum state the system is in. This type of uncertainty is the one we are used to in AI, it results from a lack of knowledge about the system. 

Mixed states are not naturally described by wave vectors. For that we need a tool called the density matrix $\rho$. For a pure state we use $\rho=\ket{\psi}\bra{\psi}$, a rank-1 matrix (or outer product), where $\bra{\psi}$ is the complex transpose of the vector $\ket{\psi}$. But for a mixed state the rank will be higher and $\rho$ can be decomposed as $\rho=\sum_k p_k \ket{\psi_k}\bra{\psi_k}$ with $\{p_k\}$ (positive) probabilities that sum to 1. Note that a unitary transformation will change the basis but not the rank (and hence will keep pure states pure): $\text{Rank}(\rho')= \text{Rank}(U\rho U^\dagger)$. In particular, time evolution will preserve rank and keep pure states pure. 

The probability of a measurement is given by the trace of the density matrix over the projector $\mA_i=\ket{\phi_i}\bra{\phi_i}$, namely $p_i=\text{Tr}(\mA_i \rho)=\text{Tr}(\ket{\phi_i}\bra{\phi_i}\ket{\psi}\bra{\psi})=|\bra{\phi_i}\ket{\psi}|^2$, i.e. Born's rule. Similar to marginalisation in classical probability theory, we can trace over degrees of freedom we are not interested in, i.e. $\rho_a=\text{Tr}_b(\rho_{ab})$. 

Further, if two operators $\mA, \mA'$ commute, they have a common eigenbasis and we can simultaneously measure them leading to a joint probability distribution $p(\lambda_\alpha, \lambda'_\beta) = \bra{\psi} \Pi_{\lambda_\alpha}\Pi_{\lambda'_\beta} \ket{\psi}$, where $\lambda_\beta'$ are the eigenvalues of $\mA'$ and $\Pi_{\lambda_\alpha}$ ($\Pi_{\lambda_\alpha'}$) projects onto the eigenspace of $\mA$ ($\mA'$).

\subsection{Review of quantum phase estimation}
\label{sec:qpe_details}
We review here the quantum phase estimation, a quantum algorithm to estimate the eigenphases of a unitary $\mU$. Suppose first that we know an eigenvector $\ket{v}$ of $\mU$ with eigenvalue
$\exp(\tfrac{2\pi i}{2^t}\varphi)$, and that $\varphi$ can be represented with $t$ bits: $\varphi = 2^{t-1} \varphi^1 + \dots + 2^{0} \varphi^t$.
Then introduce $t$ ancilla qubits in equal weight superposition of all the $2^t$ states: $H^{\otimes t}\ket{0}^{\otimes t}$,
\begin{align}
H = 
\frac{1}{\sqrt{2}}
\begin{pmatrix}
1 & 1\\
1 & -1
\end{pmatrix} 
\,,
\quad
H^{\otimes t}\ket{0}^{\otimes t}
=
\left(\frac{1}{\sqrt{2}} \ket{0} + \frac{1}{\sqrt{2}} \ket{1} \right)^{\otimes t}
=
\frac{1}{2^{t/2}}
\sum_{\ell=0}^{2^t-1}
\ket{\ell}
\,.
\end{align}
where we used the identification: $\ket{0}\equiv (1,0)^T$ and 
introduced the basis $\{ \ket{\ell} \}_{\ell=0}^{2^t-1}$ for the
ancilla qubits, i.e. $\ell = 2^{t-1} q^1 + \dots + 2^{0} q^t$. We use the ancilla's as control qubits for applying powers of $\mU$ on the input state $\ket{v}$, implementing the following unitary map:
\begin{align}
    \frac{1}{2^{t/2}}
    \sum_{\ell=0}^{2^t-1}
    \ket{\ell}
    \otimes 
    \ket{v}
    \mapsto
    \frac{1}{2^{t/2}}
    \sum_{\ell=0}^{2^t-1}
    \ket{\ell}
    \otimes 
    \mU^\ell 
    \ket{v}
    =
    \frac{1}{2^{t/2}}
    \sum_{\ell=0}^{2^t-1}
    {\rm e}^{\frac{2\pi i}{2^t} \ell\varphi}    
    \ket{\ell} \otimes \ket{v}
    \,.
\end{align}
Next, we apply the inverse Fourier transform on the ancilla register to get:
\begin{align}
    \frac{1}{2^{t}}
    \sum_{\ell,k=0}^{2^t-1}
    {\rm e}^{\frac{2\pi i \ell}{2^t} (\varphi - k )}
    \ket{k} \otimes \ket{v}
    =
    \sum_{k=0}^{2^t-1}
    \delta_{\varphi, k}
    \ket{k} \otimes \ket{v}
    =
    \ket{\varphi}
    \otimes \ket{v}
    \,.
\end{align}
The quantum complexity of this operation is linear in $t$. By linearity if the input is a generic state, $\ket{\psi}=\sum_{\alpha}\ket{v_\alpha} \braket{v_\alpha}{\psi}$, $\ket{v_\alpha}$ an eigenstate of $\mU$ with eigenvalue $\exp(\tfrac{2\pi i}{2^t}\varphi_\alpha)$, quantum phase estimation will act as:
\begin{align}
    \label{eq:qpe_psi2}
    \ket{0}^{\otimes t}
    \otimes 
    \ket{\psi}
    \mapsto
    \sum_{\alpha}
    \braket{v_\alpha}{\psi}
    \ket{\varphi_\alpha}\otimes \ket{v_\alpha} \,.
\end{align}
Here we also assumed that we can represent all the $\varphi_\alpha$'s  with $t$ bits, in which case the reduced density matrix of the ancilla state is diagonal
\begin{align}
    \label{eq:rhoanc}
    \rho_{\text{anc}}
    =
    \sum_{\alpha\beta}
    \ket{\varphi_\alpha}\bra{\varphi_\beta}
    \braket{v_\alpha}{\psi}
    \braket{\psi}{v_\beta}
    \Tr{
    \ket{v_\alpha}\bra{v_\beta}
    }
    =
    \sum_{\alpha}
    \ket{\varphi_\alpha}\bra{\varphi_\alpha}
    |\braket{v_\alpha}{\psi}|^2
    \,,
\end{align}
so that the probability of measuring $\varphi_\alpha$ is 
governed by $|\braket{v_\alpha}{\psi}|^2$. In particular, the probability that the first ancilla bit is $b$ is given by:
\begin{align}
\label{eq:p_first_bit2}
p(b)
=
\sum_\alpha |\braket{v_\alpha}{\psi}|^2 \delta(\sigma(2^{-t}\varphi_\alpha) - b)    \,,
\end{align}
where $\sigma$ is the threshold non-linearity introduced in \eqref{eq:hell} since $2^{-t}\varphi = 
2^{-1} \varphi^1 + \dots + 2^{-t} \varphi^t$ and if the first bit $\varphi^1 = 0$ then $2^{-t}\varphi < \frac{1}{2}$ and $\sigma(2^{-t}\varphi)=0$, while if 
$\varphi^1 = 1$, then 
$2^{-t}\varphi \ge \frac{1}{2}$ and $\sigma(2^{-t}\varphi)=1$.

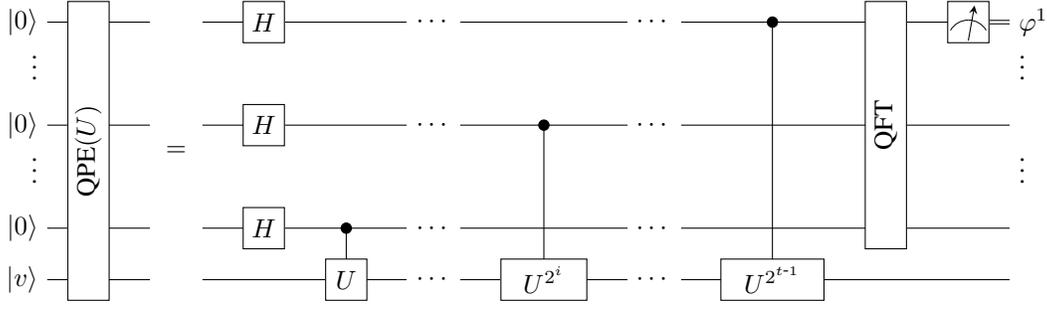
\begin{figure}[ht]
\begin{center}
\providecommand{\mynus}[1][.7]{\scalebox{#1}{-}}
\begin{tikzpicture}[scale=1.300000,x=1pt,y=1pt]
\filldraw[color=white] (0.000000, -7.500000) rectangle (280.000000, 82.500000);
\draw[color=black] (0.000000,75.000000) -- (268.000000,75.000000);
\draw[color=black] (268.000000,74.500000) -- (280.000000,74.500000);
\draw[color=black] (268.000000,75.500000) -- (280.000000,75.500000);
\draw[color=black] (0.000000,75.000000) node[left] {$\ket{0}$};
\draw[color=black] (0.000000,60.000000) node[anchor=mid east] {$\vdots$};
\draw[color=black] (0.000000,45.000000) -- (280.000000,45.000000);
\draw[color=black] (0.000000,45.000000) node[left] {$\ket{0}$};
\draw[color=black] (0.000000,30.000000) node[anchor=mid east] {$\vdots$};
\draw[color=black] (0.000000,15.000000) -- (280.000000,15.000000);
\draw[color=black] (0.000000,15.000000) node[left] {$\ket{0}$};
\draw[color=black] (0.000000,0.000000) -- (280.000000,0.000000);
\draw[color=black] (0.000000,0.000000) node[left] {$\ket{v}$};
\draw (12.000000,75.000000) -- (12.000000,0.000000);
\begin{scope}
\draw[fill=white] (12.000000, 37.500000) +(-45.000000:8.485281pt and 61.518290pt) -- +(45.000000:8.485281pt and 61.518290pt) -- +(135.000000:8.485281pt and 61.518290pt) -- +(225.000000:8.485281pt and 61.518290pt) -- cycle;
\clip (12.000000, 37.500000) +(-45.000000:8.485281pt and 61.518290pt) -- +(45.000000:8.485281pt and 61.518290pt) -- +(135.000000:8.485281pt and 61.518290pt) -- +(225.000000:8.485281pt and 61.518290pt) -- cycle;
\draw (12.000000, 37.500000) node {\rotatebox{90}{QPE$(U)$}};
\end{scope}
\draw[fill=white,color=white] (30.000000, -6.000000) rectangle (45.000000, 81.000000);
\draw (37.500000, 37.500000) node {$=$};
\begin{scope}
\draw[fill=white] (63.000000, 75.000000) +(-45.000000:8.485281pt and 8.485281pt) -- +(45.000000:8.485281pt and 8.485281pt) -- +(135.000000:8.485281pt and 8.485281pt) -- +(225.000000:8.485281pt and 8.485281pt) -- cycle;
\clip (63.000000, 75.000000) +(-45.000000:8.485281pt and 8.485281pt) -- +(45.000000:8.485281pt and 8.485281pt) -- +(135.000000:8.485281pt and 8.485281pt) -- +(225.000000:8.485281pt and 8.485281pt) -- cycle;
\draw (63.000000, 75.000000) node {$H$};
\end{scope}
\begin{scope}
\draw[fill=white] (63.000000, 45.000000) +(-45.000000:8.485281pt and 8.485281pt) -- +(45.000000:8.485281pt and 8.485281pt) -- +(135.000000:8.485281pt and 8.485281pt) -- +(225.000000:8.485281pt and 8.485281pt) -- cycle;
\clip (63.000000, 45.000000) +(-45.000000:8.485281pt and 8.485281pt) -- +(45.000000:8.485281pt and 8.485281pt) -- +(135.000000:8.485281pt and 8.485281pt) -- +(225.000000:8.485281pt and 8.485281pt) -- cycle;
\draw (63.000000, 45.000000) node {$H$};
\end{scope}
\begin{scope}
\draw[fill=white] (63.000000, 15.000000) +(-45.000000:8.485281pt and 8.485281pt) -- +(45.000000:8.485281pt and 8.485281pt) -- +(135.000000:8.485281pt and 8.485281pt) -- +(225.000000:8.485281pt and 8.485281pt) -- cycle;
\clip (63.000000, 15.000000) +(-45.000000:8.485281pt and 8.485281pt) -- +(45.000000:8.485281pt and 8.485281pt) -- +(135.000000:8.485281pt and 8.485281pt) -- +(225.000000:8.485281pt and 8.485281pt) -- cycle;
\draw (63.000000, 15.000000) node {$H$};
\end{scope}
\draw (87.000000,15.000000) -- (87.000000,0.000000);
\begin{scope}
\draw[fill=white] (87.000000, -0.000000) +(-45.000000:8.485281pt and 8.485281pt) -- +(45.000000:8.485281pt and 8.485281pt) -- +(135.000000:8.485281pt and 8.485281pt) -- +(225.000000:8.485281pt and 8.485281pt) -- cycle;
\clip (87.000000, -0.000000) +(-45.000000:8.485281pt and 8.485281pt) -- +(45.000000:8.485281pt and 8.485281pt) -- +(135.000000:8.485281pt and 8.485281pt) -- +(225.000000:8.485281pt and 8.485281pt) -- cycle;
\draw (87.000000, -0.000000) node {$U$};
\end{scope}
\filldraw (87.000000, 15.000000) circle(1.500000pt);
\draw[color=black] (112.500000, 75.000000) node [fill=white] {$\cdots$};
\draw[color=black] (112.500000, 45.000000) node [fill=white] {$\cdots$};
\draw[color=black] (112.500000, 15.000000) node [fill=white] {$\cdots$};
\draw[color=black] (112.500000, 0.000000) node [fill=white] {$\cdots$};
\draw (144.500000,45.000000) -- (144.500000,0.000000);
\begin{scope}
\draw[fill=white] (144.500000, -0.000000) +(-45.000000:17.677670pt and 8.485281pt) -- +(45.000000:17.677670pt and 8.485281pt) -- +(135.000000:17.677670pt and 8.485281pt) -- +(225.000000:17.677670pt and 8.485281pt) -- cycle;
\clip (144.500000, -0.000000) +(-45.000000:17.677670pt and 8.485281pt) -- +(45.000000:17.677670pt and 8.485281pt) -- +(135.000000:17.677670pt and 8.485281pt) -- +(225.000000:17.677670pt and 8.485281pt) -- cycle;
\draw (144.500000, -0.000000) node {$U^{2^i}$};
\end{scope}
\filldraw (144.500000, 45.000000) circle(1.500000pt);
\draw[color=black] (176.500000, 75.000000) node [fill=white] {$\cdots$};
\draw[color=black] (176.500000, 45.000000) node [fill=white] {$\cdots$};
\draw[color=black] (176.500000, 15.000000) node [fill=white] {$\cdots$};
\draw[color=black] (176.500000, 0.000000) node [fill=white] {$\cdots$};
\draw (211.000000,75.000000) -- (211.000000,0.000000);
\begin{scope}
\draw[fill=white] (211.000000, -0.000000) +(-45.000000:21.213203pt and 8.485281pt) -- +(45.000000:21.213203pt and 8.485281pt) -- +(135.000000:21.213203pt and 8.485281pt) -- +(225.000000:21.213203pt and 8.485281pt) -- cycle;
\clip (211.000000, -0.000000) +(-45.000000:21.213203pt and 8.485281pt) -- +(45.000000:21.213203pt and 8.485281pt) -- +(135.000000:21.213203pt and 8.485281pt) -- +(225.000000:21.213203pt and 8.485281pt) -- cycle;
\draw (211.000000, -0.000000) node {$U^{2^{t\mynus1}}$};
\end{scope}
\filldraw (211.000000, 75.000000) circle(1.500000pt);
\draw (244.000000,75.000000) -- (244.000000,15.000000);
\begin{scope}
\draw[fill=white] (244.000000, 45.000000) +(-45.000000:8.485281pt and 50.911688pt) -- +(45.000000:8.485281pt and 50.911688pt) -- +(135.000000:8.485281pt and 50.911688pt) -- +(225.000000:8.485281pt and 50.911688pt) -- cycle;
\clip (244.000000, 45.000000) +(-45.000000:8.485281pt and 50.911688pt) -- +(45.000000:8.485281pt and 50.911688pt) -- +(135.000000:8.485281pt and 50.911688pt) -- +(225.000000:8.485281pt and 50.911688pt) -- cycle;
\draw (244.000000, 45.000000) node {\rotatebox{90}{QFT}};
\end{scope}
\draw[fill=white] (262.000000, 69.000000) rectangle (274.000000, 81.000000);
\draw[very thin] (268.000000, 75.600000) arc (90:150:6.000000pt);
\draw[very thin] (268.000000, 75.600000) arc (90:30:6.000000pt);
\draw[->,>=stealth] (268.000000, 69.600000) -- +(80:10.392305pt);
\draw[color=black] (280.000000,75.000000) node[right] {$\varphi^1$};
\draw[color=black] (280.000000,60.000000) node[anchor=mid west] {$\vdots$};
\draw[color=black] (280.000000,30.000000) node[anchor=mid west] {$\vdots$};
\end{tikzpicture}
\end{center}
\caption{Quantum phase estimation circuit with measurement of the first ancilla qubit.}
\label{fig:qpe}
\end{figure}

Note that computing the reduced density matrix can be done by simply discarding qubits on a quantum computer, while it can be exponentially hard classically.
In practice, $t$ can to be chosen to be lower than the precision of $\varphi$ and in that case one can estimate the accuracy of the measurement, see \cite{NielsenChuang} for details.
We depict the quantum phase estimation with measurement of the first ancilla in \figref{fig:qpe}.

\section{Detailed implementation of quantum deformed neural networks}
\label{sec:details_qdnn}

We here give a detailed derivation of the formulas related to implementing quantum circuit of a layer in a quantum deformed neural network depicted in \figref{fig:circuits} (a).
The input state to that circuit is 
$\ket{0}^{\otimes t M} \otimes \ket{\psi}$, where
we recall that
\begin{align}
    \ket{\psi}
    =
    \ket{\psi}_h
    \otimes 
    \ket{\psi}_W
    \,,\quad
    \ket{\psi}_W
    =
    \bigotimes_{i=1}^N
    \bigotimes_{j=1}^M
    \left[\sqrt{q_{ji}(W_{ji}=0)} \ket{0} + \sqrt{q_{ji}(W_{ji}=1)} \ket{1}\right]
    \,.
\end{align}
We start by applying the first quantum phase estimation of $\mU_1$ which involves only the first $t$ ancillas. 
W.r.t.~\eqref{eq:qpe_psi} we identify $\ket{v_\alpha} \equiv \ket{\bm{h},\mW}_{\mD}$, $\ket{\varphi_\alpha}\equiv \ket{\varphi(\bm{h},\mW_{1,:})}$, 
to get:
\begin{align}
    &\ket{0}^{\otimes t}
    \otimes 
    \ket{\psi}
    \mapsto
    \sum_{\bm{h}\in \mathbb{B}^N}
    \sum_{\mW\in \mathbb{B}^{N M}}
    {}_{\mD}\braket{\bm{h},\mW}{\psi}
    \ket{\varphi(\bm{h},\mW_{1,:})}
    \otimes 
    \ket{\bm{h},\mW}_{\mD}
    \,.
\end{align}
Then we proceed to applying to the result the second quantum phase estimation involving $\mU_2$ and the second batch of $t$ ancilla qubits:
\begin{align}
    \sum_{\bm{h}\in \mathbb{B}^N}
    \sum_{\mW\in \mathbb{B}^{N M}}
    {}_{\mD}\braket{\bm{h},\mW}{\psi}
    \ket{0}^{\otimes t}
    \otimes
    \ket{\varphi(\bm{h},\mW_{1,:})}
    \otimes 
    \ket{\bm{h},\mW}_{\mD}
    \mapsto\\
    \sum_{\bm{h}\in \mathbb{B}^N}
    \sum_{\mW\in \mathbb{B}^{N M}}
    {}_{\mD}\braket{\bm{h},\mW}{\psi}
    \ket{\varphi(\bm{h},\mW_{2,:})}
    \otimes
    \ket{\varphi(\bm{h},\mW_{1,:})}
    \otimes 
    \ket{\bm{h},\mW}_{\mD}
    \,.
\end{align}
We repeat the procedure $M$ times to get to the final state:
\begin{align}
    \sum_{\bm{h}\in \mathbb{B}^N}
    \sum_{\mW\in \mathbb{B}^{N M}}
    {}_{\mD}\braket{\bm{h},\mW}{\psi}
    \ket{\varphi(\bm{h},\mW)}
    \otimes 
    \ket{\bm{h},\mW}_{\mD}
    \,,\quad
    \ket{\varphi(\bm{h},\mW)}
    \equiv
    \bigotimes_{j=1}^M
    \ket{\varphi(\bm{h},\mW_{j,:})}
    \,,
\end{align}
and compute the reduced density matrix of the ancilla qubits $\rho_{\text{anc}}$ which is diagonal as in \eqref{eq:rhoanc}:
\begin{align}
    \rho_{\text{anc}}
    &=
    \sum_{\bm{h}\in \mathbb{B}^N}
    \sum_{\mW\in \mathbb{B}^{N M}}
    |{}_{\mD}\braket{\bm{h},\mW}{\psi}|^2
    \ket{\varphi(\bm{h},\mW)}\bra{\varphi(\bm{h},\mW)} 
    \,.
\end{align}
Now we compute the outcome probability of a measurement of the first qubit in each of the $M$ registers of ancilla qubits. 
Recalling \eqref{eq:p_first_bit} and the fact that the first bit of an integer is the most significant bit, determining whether $2^{-t} \varphi(\bm{h},\mW_{j,:}) = (N+1)^{-1} \varphi(\bm{h},\mW_{j,:})$ is greater or smaller than $1/2$, the probability of outcome $\bm{h}'=(h'_1,\dots,h_M')$ is
\begin{align}
    \label{eq:py_D2}
    p(\bm{h}') 
    &= 
    \sum_{\bm{h}\in \mathbb{B}^N}
    \sum_{\mW\in \mathbb{B}^{N M}}
    \delta(\bm{h}' - f(\mW, \bm{h}))
    \left|\braket{\psi}{\bm{h},\mW}_{\mD}\right|^2\,,
\end{align}
where $f$ is the layer function introduced in \eqref{eq:hell}.



\section{The case of convolutional layers}
\label{sec:case_conv}

We extend here the model of \ref{sec:modifs_cl} to the convolution case, where the eigenphases we want to estimate using the quantum phase estimation are:
\begin{align}
    \varphi_{i,j,d} = 
    \sum_{k=1}^{K_1}
    \sum_{\ell=1}^{K_2}
    \sum_{c=1}^{C}
    W_{k,\ell,c,d} h_{i+k,\ell+j,c}
    \,.
\end{align}
Here the kernel $\mW$ has size $(K_1, K_2, C, C')$, where
$K_1$ ($K_2$) is the kernel along the height (width) direction, while $C$ ($C'$) is the number of input (output) channels and $d=1,\dots,C'$.
Classically, we can implement the convolution by extracting patches of size $K_1 \times K_2 \times C$ from the image and perform the dot product of the patches with a flattened kernel for each output channel. 
Moving on to the quantum implementation discussed in section \ref{sec:modifs_cl}, we recall that the input activation distribution is factorized. Therefore it is encoded in a product state, and we define patches analogously to the classical case since there is no entanglement coupling the different patches. The quantum convolutional layer can then be implemented as outlined above in the classical case, replacing the classical dot product with the quantum circuit implementing the fully connected layer of section \ref{sec:modifs_cl}. The resulting quantum layer is a translation equivariant for any choice of $\mD_j$.

\section{Details of classical simulations}
\label{sec:details_classical}

We compute here the mean and variance of \eqref{eq:mu_sigma} with the choice of \eqref{eq:D_Q_P}.
Denoting $\mB_{2i}=\mB_{i}^H, \mB_{2i+1}= \mB_{i}^W$, we have
\begin{align}
    \mK_i = \mD \frac{1}{N}\mB_{2i} \mB_{2i+1} \mD^{-1}
    =
    \mM_i^{-1}
    \frac{1}{N}
    \mB_{2i} \mB_{2i+1}
    \mM_i
    \,,\quad
    \mM_i=
    \mQ_{2i,2i+1}
    \mP_{2i-1,2i} \mP_{2i+1,2i+2}
\end{align}
so the random variable associated to $\mK_i$ will have support only on the four qubits $\{ 2i-1,2i,2i+1,2i+2\}$. Thus 
$\bra{\psi} \mK_i \mK_{i'} \ket{\psi}=
\bra{\psi} \mK_i \ket{\psi}\bra{\psi} \mK_{i'} \ket{\psi}$
for $|i-i'|>1$ and the CLT can be applied.
If we write $\ket{\psi} = \otimes_{i=0}^{2N-1}\ket{\psi_{i}}$,
and denote:
\begin{align}
    \left< \mX \right>_{i:i'}
    \equiv
    \bra{\psi_{i}} \cdots \bra{\psi_{i'}}
    \mX
    \ket{\psi_{i}} \cdots \ket{\psi_{i'}}
    \,,
\end{align}
the mean and variances are:
\begin{align}
    \label{eq:mu}
    \mu &= 
    \sum_{i=0}^{N-1} \mu_i\,,\quad
    \mu_i = 
    \begin{cases}
    \left< \mK_0 \right>_{0:2} & i=0\\
    \left< \mK_i \right>_{2i-1:2i+2} & 0<i<N-1\\
    \left< \mK_{N-1} \right>_{2N-3:2N-1} & i=N-1\\
    \end{cases}
    \\
    \label{eq:sigma_sq}
    \sigma^2 
    &=
    \sum_{ij}
    (\bra{\psi}\mK_i \mK_j \ket{\psi} - 
    \bra{\psi}\mK_i\ket{\psi}\bra{\psi}\mK_j \ket{\psi})
    \\
    &=
    2\sum_{i<j}
    (\bra{\psi}\mK_i \mK_j \ket{\psi} - 
    \bra{\psi}\mK_i\ket{\psi}\bra{\psi}\mK_j \ket{\psi})
    +
    \sum_{i=0}^{N-1}
    (\bra{\psi}\mK_i^2 \ket{\psi} - 
    \bra{\psi}\mK_i\ket{\psi}^2)    
    \\
    &=
    2 \sum_{i=0}^{N-2}
    (
    \gamma_{i,i+1}
    -
    \mu_i \mu_{i+1}
    )
    +
    \sum_{i=0}^{N-1}(\mu_i - \mu_i^2)\\
    \label{eq:gamma}
    \gamma_{i,i+1} &=
    \bra{\psi} \mK_i \mK_{i+1} \ket{\psi} =
    \begin{cases}
    \left< \mK_0\mK_1 \right>_{0:4} & i=0\\
    \left< \mK_i\mK_{i+1} \right>_{2i-1:2i+4} & 0<i<N-2\\
    \left< \mK_{N-2}\mK_{N-1} \right>_{2N-3:2N-1} & i=N-2
    \end{cases}
    \,.
\end{align}
In computing $\sigma^2$ we used $\mK_i^2 = \mK_i$.
Note that both $\mu$ and $\sigma^2$ can be computed in $O(N)$ and parallelized. Naively, computing $\left< \mK_i \right>_{2i-1:2i+2}$ and $\left< \mK_i\mK_{i+1} \right>_{2i-1:2i+4}$ takes $(d^2)^3 + d$, (the first term comes from the three layers of matrix vector products shown in \figref{fig:D} (b,c) and the second term from the final dot product) where $d$ is the dimensionality of the input state, i.e.~$d=2^4$ in the mean case and $d=2^6$ in the variance case. These constants can be further reduced by choosing an order of contractions of the tensors exploiting the local connectivities.
These times can be further greatly reduced if we assume that $\mP_{2i-1,2i}=1$, in which case $\mK_i$ is supported only on two sites and $\mK_i$, $\mK_j$ are uncorrelated for all $i,j$.

\section{Details of the experiments and additional numerical results}
\label{sec:details_experiments}

We summarize here details of the experiments discussed in section \ref{sec:exp}.
We parametrize the $4\times 4$ unitaries $\mP^j_{i,i+1}, \mQ^j_{i,i+1}$ as follows. For each $4\times 4$ unitary $\mU$ we introduce $4\times 4$ real matrices $\mA, \mB$, which are learnable. Then we compute $\mU$ as in Procedure \ref{algu}. Note that the bandPart routine effectively discards half of the parameters, halving the actual number of trainable variables, and we use this routine only for implementation convenience. We will keep that into account when counting parameters below.

\begin{algorithm} 
\caption{Parametrization of a unitary matrix}
\label{algu}
\begin{algorithmic} 
    \REQUIRE $\mA, \mB$ \COMMENT{$N\times N$ real matrices}
    \ENSURE $\mU$ 
    \STATE $\mC \leftarrow \mA + i \mB$
    \STATE $\mC \leftarrow$ bandPart($\mC$, 0, -1)
    \COMMENT{Set lower triangular part, diagonal excluded, to zero}
    \STATE $\mU \leftarrow \exp(\mC - \mC^\dagger)$
    \COMMENT{$N\times N$ unitary matrix}
\end{algorithmic}
\end{algorithm}

At the last layer of the neural network we produce a number of output probabilities equal to the number of classes, each of which is a number in $[0,1]$. We then normalize these to interpret the normalized result as probability of the input to be in a given class, fed into a cross entropy loss.

For MNIST we preprocess the data by binarizing the images and using the binary value as input bit to the quantum neural network. For Fashion MNIST such procedure would result in a great loss of texture and therefore we simply normalize the pixels to $[0,1]$ and use that value as input probability.
We train the models with the Adam optimizer and $\beta=10^{-6}$ coefficient of the variance regularization term as in \cite{pbnn}. We train for a fixed budget of $50$ epochs for MNIST Arch.~A,B,C and FashionMNIST Arch.~C, and $100$ epochs for Fashion-MNIST Arch.~A,B. We sweep over the coefficient of the $L_2$ regularization for the deformation parameters beween values $\{0, 10^{-4}\}$ and also sweep the learning rate schedule, between 
constant and equal to $0.01$ and piecewise decay from $0.01$ to $0.001$, choosing the best model among all these.
On top of the learnable parameters discussed so far, we also add bias terms to the layers in all runs.

Table \ref{tab:results2} extends the results of table \ref{tab:results}. On top of the models discussed in the main text, we also present results for models with quantum circuits constrained to be translation invariant, which have a reduced parameter count. In general, those perform less well than the unconstrained models. In a few cases, the deformed models performed slightly worse than the undeformed ones because of the more complex optimization procedure.

To understand the practical benefit of the deformations studied, we also compare the results against classical baselines with increased number of parameters to match or slightly exceed the parameter count of the deformed layers.
We start by noting that for a dense layer with $n$ inputs and $m$ outputs a classical probabilistic binary neural network baseline has $mn+m$ parameters (weights plus bias). Recalling that a $4\times 4$ unitary can be parametrized in terms of its logarithm, an anti-Hermitian matrix with $16$ real free parameters, for the quantum deformations considered we have the following ratio of deformed over classical parameters per dense layer:
\begin{align}
    \text{[PQ]}&: \frac{m  n  (1 + 16 \times 2) + m}{m n + m} \sim 1 + 16 \times 2 = 33 \\
    \text{[Q]}&: \frac{m  n  (1 + 16) + m}{m n + m} \sim 1 + 16 = 17 \\
    \text{[PQ T-inv]}&: \frac{m  n  + m 16 \times 2 + m}{m n + m} = 1 + \frac{32}{n+1}\\
    \text{[Q T-inv]}&: \frac{m  n  + m 16 + m}{m n + m} = 1+\frac{16}{n+1}
\end{align}
The $\times 2$ accounts for $\mP, \mQ$ and the 
deformation notation is explained in table \ref{tab:results2}.
When considering the number of parameters, we note that in principle the parameterization of deformed layers is redundant. Indeed the weight parameters $q_{ij}$ that make up the amplitudes of weights states $\ket{\psi_{2i-1}}$, and which correspond to the weights parameters of the classical baseline, could be incorporated into the definition of $\mP,\mQ$ since those are generic unitaries. We however count them separately to reflect the parameterization used in our experiments.

\begin{table}
\caption{
Test accuracies for MNIST and Fashion MNIST as function of parameters for various architectures.
The notation $cKsS-C$ indicates a conv2d layer with $C$ filters of size $[K,K]$ and stride $S$, and $dN$ a dense layer with $N$ output neurons.
The deformations are:
[/]: $\mP^j_{i,i+1}\!=\!\mQ^j_{i,i+1}\!=\!{\bf 1}$ (baseline \cite{pbnn});
[PQ]: $\mP^j_{i,i+1},\mQ^j_{i,i+1}$ generic;
[PQ T-inv]: $\mP^j_{i,i+1}\!=\!\mP^j,\mQ^j_{i,i+1}\!=\!\mQ^j$;
[Q]: $\mP^j_{i,i+1}\!=\!{\bf 1},\mQ^j_{i,i+1}$ generic;
[Q] T-inv: $\mP^j_{i,i+1}\!=\!{\bf 1},\mQ^j_{i,i+1}\!=\!\mQ^j$.
}
\begin{center}
\begin{tabular}{ccccc}
\hline
Architecture & Deformation & $\#$ params &  
MNIST & Fashion MNIST \\
\hline
d10
& [/] & 7850 & 91.1 & 84.2\\
d10
& [PQ] & 258730 & {\bf 94.3} & {\bf 86.8}\\
d10
& [Q] & 133290 & 91.6 & 85.1\\
d10
& [PQ T-inv] & 8170 & 91.9 & 84.3\\
d10
& [Q T-inv] & 8010 & 89.3 & 84.4\\
\hline
c3s2-8, c3s2-16, d10
& [/, /, /] & 7018 & 96.6 & 87.5\\
c3s2-8, c3s2-22, d10
& [/, /, /] & 9616 & 96.7 & 86.8 \\
c3s2-9, c3s2-21, d10
& [/, /, /] & 9382 & 96.7 & 86.7 \\
c3s2-10, c3s2-21, d10
& [/, /, /] & 9581 & 97.1 & 87.0 \\
c3s2-12, c3s2-20, d10
& [/, /, /] & 9510 & 97.1 & 86.8 \\
c3s2-8, c3s2-16, d10
& [PQ, /, /] & 9322 & {\bf 97.6} & {\bf 88.1}\\
c3s2-8, c3s2-16, d10
& [Q, /, /] & 8170 & 96.8 & 87.8\\
c3s2-8, c3s2-16, d10
& [PQ T-inv, /, /] & 7274 & 97.4 & 87.8 \\
c3s2-8, c3s2-16, d10
& [Q T-inv, /, /] & 7146 & 96.6 & 87.6\\
\hline
c3s2-32, c3s2-64, d10
& [/, /, /] & 41866 & 98.1 & 89.3 \\
c3s2-43, c3s2-68, d10
& [/, /, /] & 51304 & 98.2 & 89.4 \\
c3s2-44, c3s2-67, d10
& [/, /, /] & 51169 & {\bf 98.4} & 89.4 \\
c3s2-48, c3s2-64, d10
& [/, /, /] & 51242 & 98.3 & 89.5 \\
c3s2-32, c3s2-64, d10
& [PQ, /, /] & 51082 & 98.3 & {\bf 89.6}\\
c3s2-32, c3s2-64, d10
& [Q, /, /] & 46474 & 98.3 & 89.5\\
c3s2-32, c3s2-64, d10
& [PQ T-inv, /, /] & 42890 & 98.2 & 89.13\\
c3s2-32, c3s2-64, d10
& [Q T-inv, /, /] & 42378 & 98.3 & 89.0\\
\hline
\end{tabular}
\end{center}
\label{tab:results2}
\end{table}

\end{document}